\documentclass[twocolumn,letterpaper]{article}
\usepackage[hmargin=1.6cm,vmargin=2cm,centering]{geometry}

    \usepackage[cmex10]{amsmath}
    \usepackage{amsthm}
    \usepackage{url,xspace,array}
    \usepackage{cite}
    \usepackage[all]{xypic}
    \usepackage[utf8]{inputenc}

    \usepackage{ssect}
    \usepackage{macros}

    \newcommand{\myparagraph}[1]{\vspace{.15cm}\noindent \textbf{#1}}

\begin{document}
    \title{An Epistemic Approach to Coercion-Resistance\\ for Electronic
           Voting Protocols\thanks{%
            This paper is an extended version of \cite{KuestersTruderung-SP-2009}.
           }}
    \author{
        Ralf K\"usters and Tomasz Truderung \\[1ex]
            \normalsize
            \normalsize University of Trier\\ 
            \normalsize Germany\\
            \normalsize Email: \{kuesters,truderun\}@uni-trier.de
    }

    \maketitle
    
    \begin{abstract}\small
      Coercion resistance is an important and one of the
      most intricate security requirements of electronic
      voting protocols. Several definitions of coercion
      resistance have been proposed in the literature,
      including definitions based on symbolic models.
      However, existing definitions in such models are
      rather restricted in their scope and quite complex.

      In this paper, we therefore propose a new definition
      of coercion resistance in a symbolic setting, based
      on an epistemic approach. Our definition is
      relatively simple and intuitive. It allows for a
      fine-grained formulation of coercion resistance and
      can be stated independently of a specific, symbolic
      protocol and adversary model. As a proof of concept,
      we apply our definition to three voting
      protocols. In particular, we carry out the first
      rigorous analysis of the recently proposed Civitas
      system.  We precisely identify those conditions under
      which this system guarantees coercion resistance or
      fails to be coercion resistant. We also analyze
      protocols proposed by Lee et al. and Okamoto.
    \end{abstract}
       
    \section{Introduction }

Coercion resistance is one of the most important and
intricate security requirements of voting protocols
\cite{JuelsCatalanoJakobsson-WPES-2005,Okamoto-SPW-1997}.
Intuitively, a voting protocol is coercion resistant if it
prevents voter coercion and vote buying. In other words, a
coercer should not be able to influence the behavior of a
voter. A notion closely related to coercion resistance, but
somewhat weaker is receipt freeness, first proposed in
\cite{BenalohTuinstra-STOC-1994}.

Most voting schemes and systems that aim to achieve
coercion resistance or receipt freeness come without a
rigorous security proof.  Maybe not surprisingly, some of
these protocols have been found to be flawed (see, e.g.,
discussions in \cite{MoranNaor-CCS-2007} and
\cite{HirtSako-EUROCRYPT-2000}). The lack of proofs is
partly due to the fact that only recently first formal
definitions of coercion resistance and receipt freeness
have been proposed in the literature, both based on
cryptographic and symbolic models
\cite{JuelsCatalanoJakobsson-WPES-2005,MoranNaor-CRYPTO-2006,Chevallier-MamesFouquePointchevalSternTraore-WOTE-2006,JonkerPieters-WOTE-2006,JonkerVink-ISC-2006,DelauneKremerRyan-CSFW-2006,BackesHritcuMaffei-CSF-2008}.
With ``cryptographic models'' we mean models in which
messages are modeled as bit strings and adversaries are
probabilistic polynomial time Turing machines. In contrast,
symbolic models take a more abstract view on cryptography.
In this paper, our focus will be on symbolic models. While
security guarantees in cryptographic models are typically
stronger than in symbolic models, security proofs in
cryptographic models are usually very involved, and as a
result, often omitted or only sketched. For electronic
voting protocols, which are among the most complex security
protocols, this is even more so (see, e.g.,
\cite{BenalohTuinstra-STOC-1994,ClarksonChongMyers-SP-2008,JuelsCatalanoJakobsson-WPES-2005,HirtSako-EUROCRYPT-2000,SakoKilian-EUROCRYPT-1995,Okamoto-SPW-1997,LeeBoydDawsonKimYangYoo-ICISC-2003}).
Conversely, security proofs in symbolic models are easier
to carry out and they are more amenable to tool support.
Research on security protocol analysis has demonstrated
that, while not all, but many attacks on security protocols
can be uncovered and prevented by means of symbolic
protocol analysis (see, e.g.,
\cite{MitchellShmatikov+-USENIX-1998,CervesatoJaggardScedrovTsayWalstad-WITS-2006,BellaMassacciPaulson-IJIS-2005,BhargavanFournetGordonTse-CSFW-2006,ArmandoBasinetal-CAV-2005,Blanchet-CSFW-2001,KuestersTruderung-CCS-2008}).
In some cases, security guarantees established in symbolic
models even imply security in cryptographic models (see,
e.g.,
\cite{AbadiRogaway-IFIPTCS-2000,MicciancioWarinschi-TCC-2004,CortierKremerKuestersWarinschi-FSTTCS-2006}).
Hence, symbolic models certainly have their merits for
security protocol analysis, including the analysis of
voting protocols.

However, the definitions of coercion resistance in symbolic
models proposed in the literature thus far are rather
restricted in scope, yet quite complex and not always
intuitive (see Section~\ref{sec:relatedwork} for a detailed
discussion).

\myparagraph{Contribution of this paper.}  One of the main
contributions of this paper is to provide a general, yet
intuitive and simple definition of coercion resistance. Our
definition follows an epistemic approach. It is formulated
in a model-independent way. In particular, it can be
instantiated by different symbolic models. While the focus of
this work is on voting protocols, our definition may be
applicable beyond this domain. 

In order to analyze concrete voting protocols, we
instantiate our framework by a rather standard symbolic
model. Within our model, we prove several general
statements, which underline the adequacy of our model and
which have not been proven in other symbolic models. Among
others, we show that coercion resistance w.r.t.~a single
coerced voter implies coercion resistance w.r.t.~multiple
coerced voters.

As a proof of concept, we analyze coercion resistance of
three voting protocols: the recently proposed voting system
Civitas \cite{ClarksonChongMyers-SP-2008}, a voting
protocol by Lee et
al.~\cite{LeeBoydDawsonKimYangYoo-ICISC-2003} and one by
Okamoto \cite{Okamoto-SPW-1997}. As to the best of our
knowledge, Civitas and the scheme by Okamoto have not been
rigorously analyzed before. Our modeling, in particular of
Civitas, is quite detailed and goes beyond the level of
detail considered in other works. For example, for Civitas
we model dishonest authorities and the zero-knowledge
proofs authorities have to provide to prove their
compliance with the protocol. We precisely identify those
conditions under which coercion resistance is guaranteed
and point out situations in which the protocols do not
provide coercion resistance, thereby relativizing previous
claims and providing new insights into and improvements of
the protocols. The analyzes of the example protocols
illustrate that our definition of coercion resistance
allows to specify various degrees of coercion resistance in
a fine-grained way. Without this flexibility of our
definition, no reasonable statement about the coercion
resistance of voting protocols would be possible as every
protocol builds on its own assumptions and provides
specific security guarantees.

\myparagraph{Structure of this paper.} In the following
section, we present our definition of coercion resistance.
A concrete instantiation of this definition is provided in
Section~\ref{sec:instantiationofgeneralframework}, with
general properties given in
Section~\ref{sec:generalproperties}. The analyzes of the
three mentioned voting protocols are then presented in
Sections~\ref{sec:civitas}, \ref{sec:leeprotocol}, and
Appendix \ref{sec:okamoto}.  Related work is discussed in
Section~\ref{sec:relatedwork}.  We conclude in
Section~\ref{sec:conclusion}. More details and proofs can
be found in the appendix.


\section{Defining Coercion Resistance}\label{sec:definingcoercionresistance}

In this section, we present our definition of coercion
resistance in an epistemic framework, independent of a
specific, symbolic protocol or adversary model. A concrete
instantiation will be considered in
Section~\ref{sec:instantiationofgeneralframework}.

Our definition of coercion resistance is based on what we
call a coercion system. A coercion system will be induced
by a voting protocol (see
Section~\ref{sec:instantiationofgeneralframework}). It
emphasizes in an abstract way those parts relevant for
defining coercion resistance, without the need to consider
details of a protocol and adversary model. More intuition
is provided following the next definition.

\begin{definition}
  A \emph{coercion system} is a tuple $S=(R,V,C,E,r,\sim)$,
  where $R$ is a set of runs, $V$, $C$, and $E$ are sets of
  possible programs of coerced voters, the coercer, and the
  environment, respectively, $r$ is a mapping which assigns
  a set $r(v,c,e)\subseteq R$ of \emph{runs induced by
    $(v,c,e)$} to each tuple $(v,c,e)\in V\times C\times E$, and $\sim$
  is an equivalence relation on the set $R$, which
  determines the view of a coercer on a run.
\end{definition}
A coercion system determines the possible behaviors of
coerced voters, the coercer, and the environment. The
environment is the part of the system controlled neither by
the coercer nor by the coerced voter.  The environment
typically describes the possible behaviors of honest
entities, such as honest voters and authorities; dishonest
voters and authorities will be subsumed by the coercer. The
programs carried out by these honest entities will be
determined by the voting protocol under consideration.
However, the environment typically does not fix up front
how and if certain honest voters vote. It may also leave
open the number of voters as well as how many of them and
which voters are honest or dishonest. The set $r(v,c,e)$
describes the possible runs obtained when the programs $v$,
$c$, and $e$ of the coerced voter, the coercer, and the
environment, respectively, run together. A run is typically
a sequence of configurations induced by the interaction of
$v$, $c$, and $e$. However, for a general definition of
coercion resistance it is not necessary to fix such details
at this point. The reason that we do not define $r(v,c,e)$
to be a single run is that a run of $v$, $c$, and $e$ might
involve some non-deterministic choices, e.g.,
non-deterministic scheduling of messages.  The equivalence
relation $\sim$ defines the view of the coercer. The
intuition is that if two runs $\rho$ and $\rho'$ are
equivalent w.r.t.~$\sim$, i.e., $\rho\sim \rho'$, then the
coercer has the same view in both runs. In other words,
these runs look the same from the coercer's point of view.

We can now turn to the definition of coercion resistance.
For the following discussion, we concentrate on the case
that only a single voter is coerced. The case of
multi-voter coercion resistance is discussed later.

Given a coercion system $S=(R,V,C,E,r,\sim)$, the idea
behind our definition of coercion resistance is as follows:

Our definition assumes that the coerced voter has a certain
goal $\gamma$ that he/she would try to achieve in absence of
coercion.  Formally, $\gamma$ is a subset of $R$, the set of
runs of $S$.  If, for example, $\gamma$ is supposed to
express that the coerced voter wants to vote for a certain
candidate, then $\gamma$ would contain all runs in which the
coerced voter voted for this candidate and this vote is
in fact counted.  Jumping ahead, as we will see in the
analysis of concrete protocols, often such a goal cannot be
achieved. This is, for example, the case if ballots are sent
over an unreliable channel or an election authority misbehaves
in an observable way and as a result the election process is
stopped. A more realistic goal $\gamma$ would then be that the
coerced voter successfully votes for a certain candidate,
provided the voters ballot is delivered in time and the
election authority did not misbehave in an observable way.

Now, in the definition of coercion resistance we imagine
that the coercer provides the coerced voter with a program
$v\in V$ (the \emph{coercion strategy}), which the coercer
wants the coerced voter to run, instead of the program the
coerced voter would carry out when following the voting
protocol. The program $v$ might determine the candidate for
which the coercer wants the coerced voter to vote for or
might dictate the coerced voter not to vote (abstention
attack).  The choice of the candidate or whether or not the
coerced voter should abstain from voting might even depend
on the course of the election process and the information
that the coercer has gathered thus far.  Such information
can be gathered by the program $v$ or might be given to the
program by the coercer; in the most general setting, one
assumes that the coercer can freely communicate with the
program $v$, and by this, further influence and control the
behavior of the coerced voter.  Rather than directly
manipulating the outcome of the election, the purpose of
$v$ might as well be to merely test whether the coerced
voter follows the prescribed program $v$; for example, to
find out whether this voter is ``reliable'', and hence, is
a good candidate for coercion in later elections. This
illustrates that the intentions of the coercer are manifold
and hard to predict.  The set $V$ should therefore contain
all programs that a coercer could possibly give to a
coerced voter.  However, as shown in
Section~\ref{sec:dummytheorem}, in a concrete communication
model, it often suffices to consider just one program that
simply forwards all messages from/to the coercer.
Nevertheless, taking the set $V$ into account only makes
our definition more flexible since different classes of
coercion strategies can be specified.

Our definition of coercion resistance requires that for all
$v\in V$, there exists a program $v'\in V$, the
\emph{counter strategy}, that the coerced voter can run
instead of $v$, such that (i) the voter always achieves
his/her own goal $\gamma$ by running $v'$ and (ii) the
coercer does not know whether the coerced voter run $v$ or
$v'$. In other words, in every run in which the coerced
voter run $v$, the coercer thinks, given his/her view of
the run, that it is possible that the coerced voter run
$v'$. Conversely, in every run in which the coerced voter
run $v'$, the coercer thinks that it is possible that the
coerced voter run $v$. So, the coercer cannot know whether
the coerced voter followed the coercer's instructions
(i.e., run $v$) or just tried to achieve his/her own goal
(by running $v'$). If in some situations the coercer knew
that the coerced voter run either $v$ or $v'$, then the
voter could be influenced: The coercer could give positive
and/or negative incentives for running $v$/$v'$,
e.g., by offering money and/or threatening the coerced
voter.

The above leads to the following definition.  The meaning
of $\alpha$ is explained below.

\begin{definition}\label{def:cr}
  Let $S=(R,V,C,E,r,\sim)$ be a coercion system and
  $\alpha,\gamma\subseteq R$. The system $S$ is
  \emph{coercion resistant in $\alpha$ w.r.t.\ $\gamma$},
  if for each $v\in V$ there exists $v'\in V$ such that the
  following conditions are satisfied.
  \begin{ienum}
  \item For every $c\in C$, $e\in E$, and $\rho \in
    r(v,c,e)\cap\alpha$, there exists $e'\in E$ and $\rho'
    \in r(v',c,e')$ such that $\rho \sim \rho'$.
  \item For every $c\in C$, $e\in E$, and $\rho \in
    r(v',c,e)\cap\alpha$, there exists $e'\in E$ and $\rho'
    \in r(v,c,e')$ such that $\rho \sim \rho'$.
  \item For every $c\in C$ and $e\in E$, we have 
    $r(v',c,e) \subseteq \gamma$.
  \end{ienum}
\end{definition}
Condition (iii) in the above definition directly captures
that if the coerced voter runs the counter strategy $v'$,
then independently of the actions of the coercer $c$ and
the environment $e$, the coerced voter achieves his/her
goal. To explain the conditions (i) and (ii), let us first
ignore the set $\alpha$.  Then (i) says that, for every run
$\rho$ in which the coerced voter carries out $v$, there
exists another run $\rho'$ in which the coerced voter
carries out $v'$ such that the view of the coercer, who
runs $c$ in both runs, is the same. In other words, even
though the coerced voter carried out $v$, from the
coercer's point of view it is possible that the coerced
voter carried out $v'$. The programs $e$ and $e'$ in (i)
might, for example, differ in the way honest voters voted.
So even though the coerced voter might not have voted in
the way intended by the coercer, the coercer can not tell
from the outcome of the election, as the coercer does not
have complete knowledge about how everybody voted.
Analogously, condition (ii) says that in every run in which
the coerced voter run $v'$, the coercer thinks that it is
possible that the coerced voter run $v$. Altogether (i) and
(ii) say that the coercer never knows whether the coerced
voter run $v$ or $v'$.

Now, let us discuss the purpose of $\alpha$. The intuition
is that $\alpha$ describes a property of the environment
(which, as mentioned, includes the honest voters) in terms
of a set of runs that satisfy this property.  The set
$\alpha$ typically includes almost all runs of the system,
except for those that are unlikely to happen and would
reveal to the coercer that the coerced voter is following
$v$ or $v'$.  For example, $\alpha$ would typically not
contain a run, say $\rho$, in which a certain candidate,
say $a$, does not get any vote from the honest voters.
Indeed, to obtain a successful counter strategy, it is
necessary to exclude such a run: Assume that the coercer
wants the coerced voter to vote for $a$ (hence, an
appropriate $v$ is given by the coercer to the coerced
voter). Also assume that the goal $\gamma$ of the coerced
voter is to vote for a different candidate, say $b$. Then
in the run $\rho$ from above, if the coerced voter ran the
counter strategy $v'$, the coercer would easily detect this
fact: If after the election the coercer observes that there is
no vote for $a$, the coercer can be sure that the coerced
voter was not following the coercion strategy~$v$.  In other
words, in Definition~\ref{def:cr}, if $v'$ satisfies (iii),
then (ii) cannot be satisfied, unless by $\alpha$ runs such as
$\rho$ are excluded. This example shows that without taking an
appropriate $\alpha$ into account, Definition~\ref{def:cr}
would be too strong in almost all realistic settings.

The example protocols analyzed in
Sections~\ref{sec:civitas}, \ref{sec:leeprotocol}, and
Appendix \ref{sec:okamoto} will further illustrate the
usefulness and necessity of the parameters $\alpha$ and
$\gamma$ of our definition of coercion resistance. These
parameters allow to precisely capture under what conditions
a protocol is coercion resistant, making for a quite
fine-grained and general notion of coercion resistance.

Definition~\ref{def:cr} only stipulates the existence of a
counter strategy $v'$, given a coercion strategy $v$.
However, it might in general not be easy to come up with
$v'$ given $v$. Fortunately, as already mentioned above, we
can show that it is often suffices to come up with a
counter strategy only for what we call a dummy coercion
strategy, which merely forwards messages to/from the
coercer. Given such a counter strategy, one can, in a
generic way, construct a counter strategy for any given
coercion strategy (see Section~\ref{sec:dummytheorem}). We
believe that the construction of a counter strategy from a
(dummy) coercion strategy should be part of the protocol
specification, so that a voter knows how to defend against
coercion (see also \cite{MoranNaor-CRYPTO-2006}).

We note that Definition~\ref{def:cr} captures coercion
resistance in a possibilistic way. We do not consider
probabilities. While Definition~\ref{def:cr} requires that
from the coercer's point of view it is always possible that
the coerced voter run $v$, say, the definition does not
talk about the probability for this to be the case. If this
probability were low, the coercer could tend to believe
that the coerced voter run $v'$. We leave a
probabilistic/cryptographic version of our definition as
future work. The analysis carried out in this work for the
three voting protocols shows that already in a
possibilistic setting non-trivial security guarantees can
be proved and subtle vulnerabilities can be uncovered.

While in Definition~\ref{def:cr} only one goal of the
coerced voter is considered, a protocol should of course be
coercion resistant no matter what goal the coerced voter
would like to achieve; for example, no matter which
candidate the coerced voter would like to vote for. This is
captured by the following generalization of
Definition~\ref{def:cr}.

\begin{definition}
  Let $S=(R,V,C,E,r,\sim)$ be a coercion system and
  $\Gamma$ be a set of goals, i.e.~$\Gamma$ is a set of
  subsets of $R$.  Then $S$ is \emph{coercion resistant in
    $\alpha$ w.r.t.\ $\Gamma$}, if $S$ is
  coercion resistant in $\alpha$ w.r.t.\ $\gamma$, for each
  $\gamma\in\Gamma$.
\end{definition}

\myparagraph{Multi-voter coercion.} So far, we had in mind
that $v$ and $v'$ stand for programs carried out by a
single coerced voter. Nevertheless, we can just as well
think of $v$ and $v'$ as tuples of programs carried out by
multiple coerced voters, where the tuples may be of varying
length, depending on how many voters are coerced. In other
words, our definition of coercion resistance directly
carries over to the case of \emph{multi-voter coercion
  resistance}, where multiple voters are coerced at the
same time. However, the requirement ``for all $v$ there
exists a $v'$ such that \ldots'' in the definition of coercion
resistance then only means that a coerced voter can pick a
counter strategy depending on all the programs in $v$. This
is too weak. A coerced voter should be able to pick his/her
counter strategy independently of other coerced voters; a
coerced voter may in general not know who else is coerced
and with whom he/she can (safely) collaborate.  Therefore,
for multi-voter coercion resistance, we replace the
requirement ``for all $v$ there exists a $v'$ such that
\ldots'' by ``there exists a function $f$ which maps a
coercion strategy for one voter to a counter strategy for
one voter such that, for every tuple $v$ of programs,
$v'=f(v)$ is a counter strategy such that \ldots'', where
$f(v)$ means that $f$ is applied to every single program in
the tuple $v$.

In Section~\ref{sec:multivotercoercion} we show that (a
slight extension of) coercion resistance w.r.t.~a single
coerced voter implies multi-voter coercion resistance. So,
to obtain multi-voter coercion resistance it suffices to
consider the case of a single coerced voter.


    \section{A Concrete Protocol and Adversary Model}\label{sec:instantiationofgeneralframework}

In this section, we instantiate the framework presented in
the previous section by a concrete protocol and adversary
model. Several instantiations are possible, including, for
example, one based on I/O automata or process calculus. For
the sake of brevity, we pick a quite abstract one, in which
computations are described by certain functions, called
atomic processes. However, the results presented in the
subsequent sections also carry over to other models. We
note that these sections should be intelligible without the
concrete protocol and adversary model presented in this
section.

\subsection{Terms and messages} 

Let $\Sigma$ be some
signature for cryptographic primitives (including a
possibly infinite set of constants for representing
participant names, etc.), $X=\{x_1,x_2,\dots\}$ be a set of
variables, and $\nonces$ be an infinite set of
\emph{nonces}, where the sets $\Sigma$, $X$, and $\nonces$
are pairwise disjoint.  For $N\subseteq\nonces$, the set
$T_N$ of \emph{terms} over $\Sigma\cup N$ and $X$ is
defined as usual. Ground terms, i.e., terms without
variables, represent messages.  We assume some fixed
equational theory associated with $\Sigma$ and denote by
$\equiv$ the congruence relation on terms induced by this
theory. The exact definition of $\Sigma$ and the equational
theory will depend on the cryptographic primitives used in
the voting protocol under consideration. For the voting
protocols we analyze in Section~\ref{sec:civitas},
\ref{sec:leeprotocol}, and Appendix \ref{sec:okamoto} quite
involved signatures and equational theories will be
considered, which, among others, allow to model homomorphic
encryption and various kinds of zero knowledge proofs
(designated-verifier reencryption proofs, distributed
plaintext equivalence tests, etc.). A simple example
of a signature $\Sigma_{ex}$ and its associated equational
theory is provided in
Figure~\ref{fig:examplesigmaequations}. A term of the form
$\sig{m}{k}$ represents a message $m$ signed using the
(private) key $k$. Checking validity of such a signature is
modeled by equation \eqref{eq:exchecksignature}. The
fact that signatures do not necessarily hide the signed
message is taken care of by equation
\eqref{eq:exextractmessagesig}. A term of the form $\parenc
x{\pub(k)}r$ represents the ciphertext obtained by
encrypting $x$ under the public key $\pub(k)$ using
randomness $r$. Decryption of such a term using the
corresponding private key $k$ is modeled by equation
\eqref{eq:exdecrypt}. A term of the form $\an{x,y}$ models
the pairing of terms $x$ and $y$. The components $x$
and $y$ of $\an{x,y}$ can be extracted by applying the
operators $\first{\cdot}$ and $\second{\cdot}$,
respectively, as modeled by the equations
\eqref{eq:expair}. Let $\equiv_{ex}$ denote the congruence
relation induced by the equational theory in
Figure~\ref{fig:examplesigmaequations}, then we have that $\dec{\parenc
  a{\pub(k)}r}{\first{\an{k,b}}}\equiv_{ex} a$.

\begin{figure}[t]\centering
    \begin{align}
    \label{eq:exchecksignature}        
        \checksig{\sig{m}{k}}{\pub(k)} &= \True \\[.4ex]
    \label{eq:exextractmessagesig}        
        \unsig{\sig{m}{k}} &= m \\[.4ex]
    \label{eq:exdecrypt}        
        \dec{\parenc x{\pub(k)}r}{k} &= x \\[.4ex]
    \label{eq:expair}      
        \first{\an{x,y}} = x,  \qquad  \second{\an{x,y}} &= y 
    \end{align}
    \caption{The equational theory associated with the
      signature
      $\Sigma_{ex}=\{\sig{\cdot}{\cdot},\an{\cdot,\cdot},\parenc{\cdot}{\cdot}{\cdot},\True,\checksig{\cdot}{\cdot},\unsig{\cdot},$
      $\first{\cdot},\second{\cdot}\}$.\label{fig:examplesigmaequations}.}
\end{figure}

\subsection{Event sequences and views} 

Let $\channels$ be a set of \emph{channels} (\emph{channel
names}). An \emph{input/output event} is of the form $(c:m)$ and
$(\bar c:m)$, respectively, for $c\in \channels$ and a message
$m$ (note that $\bar c\notin\channels$). A finite or infinite
sequence of events is called an \emph{event sequence}. For an
event sequence $\rho= (c_1:m_1),(c_2:m_2), \ldots$ of input
events, we denote by $\chseq(\rho)$ the sequence
$c_1,c_2,\ldots$ of channels. For $C\subseteq\channels$, we
denote by $\rho_{|C}$ the subsequence of $\rho$ containing only
the events $(c:m)$ with $c\in C$.  

Let $\tau\in T_N$ be a term.  Then, with $\rho$ as above, we denote by
$\tau[\rho]$ the message $\tau[m_1/x_1, m_2/x_2,\dots]$, where $x_i$ is
replaced by $m_i$. (Recall that the set of variables is
$X=\{x_1,x_2,\dots\}$.) For example, assume that
$\tau_{ex}=\dec{x_1}{\first{x_2}}$ and
$\rho_{ex}=(c_1:\parenc a{\pub(k)}r), (c_2:\an{k,b})$. Then
$\tau_{ex}[\rho_{ex}]=\dec{\parenc
  a{\pub(k)}r}{\first{\an{k,b}}}\equiv_{ex} a$. 

Borrowing the notion of static equivalence from
\cite{AbadiFournet-POPL-2001}, we call two event sequences
$\rho$ and $\rho'$ of input events \emph{statically
  equivalent w.r.t.~a set $C\subseteq \channels$ of
  channels and a set $N\subseteq \nonces$ of nonces},
written $\rho \equiv^C_N \rho'$, if (i) $\chseq(\rho_{|C})
= \chseq(\rho'_{|C})$ and (ii) for every $\tau_1,\tau_2 \in
T_N$ we have that $\tau_1[\rho_{|C}] \equiv
\tau_2[\rho_{|C}]$ iff $\tau_1[\rho'_{|C}] \equiv
\tau_2[\rho'_{|C}]$.  Intuitively, a party listening on
channels $C$ and a priori knowing the nonces in $N$, cannot
distinguish between the inputs received according to $\rho$
and those received according to $\rho'$. We call the
equivalence class of $\rho$ w.r.t.\ $\equiv^C_N$, the
\emph{$(C,N)$-view} on $\rho$. For example, if $k$, $k'$,
$a$, and $b$ are different constants, $r$ and $r'$ are nonces,
$C=\{c_1,c_2\}$, and $N=\emptyset$, then it is easy to see that
$\rho^1_{ex}=(c_1:\parenc a{\pub(k)}r), (c_2:\an{k',b}),(c_3:k)$
and $\rho^2_{ex}=(c_1:\parenc b{\pub(k)}{r'}),
(c_2:\an{k',b})$ yield the same $(C,N)$-view
w.r.t.~$\equiv_{ex}$.

\subsection{Processes} 

Processes are built from atomic 
processes.  An atomic process is basically a function that
given a sequence of input events (representing the history
so far) produces a sequences of output events. We require
that an atomic process behaves the same on
inputs on which it has the same view. Formally, atomic
processes are defined as follows.

\begin{definition}
    An \emph{atomic process} is a tuple $p = (I, O, N, f)$ where
    \begin{ienum}
    \item $I,O\subseteq \channels$ are finite sets of
      \emph{input} and \emph{output} channels, respectively,
    \item $N\subseteq\nonces$ is a set of \emph{nonces used by
        $p$},
    \item $f$ is a mapping which assigns a sequence $f(U) =
      (c_1:\tau_1)\cdots(c_n:\tau_n)$ with $c_i\in O$ and
      $\tau_i\in T_N$ to each $(I,N)$-view~$U$.
    \end{ienum}
    We refer to $I$, $O$ and $N$ by $\inp p$, $\outp p$, and
    $\nonc p$, respectively. We note that the sets $\inp p$ and
    $\outp p$ do not have to be disjoint (which means that $p$
    can send messages to itself).
\end{definition}
We note that (iii) guarantees that $p$ performs the same
computation on event sequences that are equivalent
according to $\equiv^I_N$, and hence, on which $p$ has the
same view. This is why $f$ is defined on $(I,N)$-views
rather than on sequences of input events.

For an event sequence $\rho$, we write $p(\rho)$ for the
output produced by $p$ on input $\rho$. This output is
$(c_1:\tau_1[\rho'])\cdots(c_n:\tau_n[\rho'])$, where
$\rho' = \rho_{|I}$ and $(c_1:\tau_1)\cdots(c_n:\tau_n) =
f(U)$ for the equivalence class $U$ of $\rho'$
w.r.t.~$\equiv^I_N$. For example, let $I=\{c_1,c_2\}$,
$N=\emptyset$, $U$ be the equivalence class of
$\rho^1_{ex}$, and assume that
$f(U)=(c_4:\an{x_1,\first{x_2}})$. Then,
$p(\rho^1_{ex})=(c_4:\an{\parenc
  a{\pub(k)}r,\first{\an{k',b}}})$, which modulo
$\equiv_{ex}$ can be equivalently written as
$(c_4:\an{\parenc a{\pub(k)}r,k'})$ and
$p(\rho^2_{ex})=(c_4:\an{\parenc
  b{\pub(k)}{r'},\first{\an{k',b}}})$, which modulo
$\equiv_{ex}$ can be equivalently written as
$(c_4:\an{\parenc b{\pub(k)}{r'},k'})$. Note that since
$\rho^1_{ex}$ and $\rho^2_{ex}$ yield the same $(I,N)$-view
w.r.t.~$\equiv_{ex}$, $p$ performs the same transformation
on $\rho^1_{ex}$ and $\rho^2_{ex}$.

For atomic processes $p$ and $p'$, we write $p \simeq p'$, if
$p$ and $p'$ perform the same computation up to renaming of
nonces. Formally, for atomic processes $p = (I, O, N, f)$, $p'=
(I, O, N', f')$, we write $p \simeq p'$, if there exists a
bijection $h:N \rightarrow N'$ such that $h(f(U)) = f'(h(U))$
for every $(I,N)$-view $U$.  This is extended to processes (see
below) in the obvious way.

A \emph{process} $P$ is a finite set of atomic processes
with disjoint sets of input channels and sets of nonces,
i.e., $\inp p\cap\inp{p'}=\emptyset$ and $\nonc
p\cap\nonc{p'}=\emptyset$, for distinct $p,p'\in P$.
The set of input/output channels and the set of nonces of
$P$ is $\inp P = \bigcup_{p\in P}\inp p$, $\outp P =
\bigcup_{p\in P}\outp p$, and $\nonc P = \bigcup_{p\in
  P}\nonc p$, respectively.  We say that $P$ is a
\emph{process over $(I,O,N)$}, if $\inp P\subseteq I$,
$\outp P\subseteq O$, and $\nonc P\subseteq N$. By
$\Proc(I,O)$ we denote the set of all processes over
$(I,O,N)$, for some $N\subseteq\nonces$. 

For a finite event sequence $\rho$ with the last event of the
form $(c:m)$, we write $P(\rho)$ for $p(\rho)$, where $p$ is the
(unique) element of $P$ such that $c\in\inp p$ (if such a $p$
does not exists, then $P(\rho)$ is undefined).

Given a process $P$ and a finite sequence $s_0$ of output
events over $\outp P$, a \emph{run $\rho$ of a process $P$
  initiated by $s_0$} is a finite or infinite sequence of
input and output events which evolves from $s_0$ in a
natural way: An output event is chosen
non-deterministically (initial from $s_0$). Once an output
event has been chosen, it will not be chosen anymore later on.
By definition of processes, there exists at most one atomic
process, say $p$, in $P$ with an input channel
corresponding to the output event. Now, $p$ (if any) is
given the input event corresponding to the chosen output
event, along with all previous input events on channels of
$p$. Then, $p$ produces a sequence of output events as
described above. Now, from these or older output events an
output event is chosen non-deterministically, and the
computation continues as before. 
The notion of a run is formally defined below.

\begin{definition}
    Let $P$ be a process and $s_0$ be a finite sequence of events.
    A \emph{run induced by $P$ initiated by $s_0$} is a
    sequence of events $s = e_1 e_2 \dots$ such that 
    \begin{ienum}
    \item
        $s$ begins with $s_0$,
      \item There exists a bijective function $f$ from
        non-negative integers to non-negative integers
        such that for each $i$, if $e_i$ is an input
        event $(c:m)$, then $f(i)<i$ and $e_{f(i)} =
        (\bar c:m)$, and, moreover, for each $i<j$ with
        $e_i = (c:m)$ and $e_j = (c:m')$ we have
        $f(i)<f(j)$,
    \item
       If $\rho$ can be splitted into $\rho_1$ $\rho_2$
       $\rho_3$, where $\rho_1$ ends with an input event, $\rho_2$
       contains output events only, and $\rho_3$ either is empty
       or begins with an input event, then $\rho_2 = P(\rho_1)$.
    \end{ienum}
\end{definition}

    An run is \emph{fair} if it is finite or, in case
    it is infinite, each message sent is eventually
    delivered (i.e. for each output event $e_i$ there
    exists $j$ such that $f(j)=i$).

A run is finite if all
output events were chosen at some point and there is no new
output event left that has not yet been chosen; otherwise a
run is infinite.  We emphasize that $s_0$ can induce many
runs, due to the non-deterministic delivery of messages.

We call two processes $P$ and $P'$ \emph{non-conflicting}
if $\inp P\cap\inp{P'}=\emptyset$ and $\nonc P\cap\nonc{P'}
= \emptyset$. In this case, we will write $P_1 \parallel
P_2$ instead of $P_1 \cup P_2$.

If $P\subseteq P'$, we call process $P$ a \emph{subprocess}
of process $P'$. For such a $P$, we define an equivalence
relation $\equiv_P$ on runs induced by $P'$ as follows:
$\rho_1 \equiv_P \rho_2$ iff $\rho_1 \equiv^{\inp P}_{\nonc
  P}\rho_2$. Hence, $\rho_1 \equiv_P \rho_2$ means that
from the point of view of $P$, the runs $\rho_1$ and
$\rho_2$ look the same. In particular, $P$ behaves the same
on these runs.

\subsection{Protocols and Their Induced Coercion Systems}  

\begin{definition}\label{def:protocol}
    A \emph{protocol} is a tuple $S =
    (A, \vin, \vout, s_0, P)$, where 
    \begin{ienum}
    \item
        $A$ is a finite set of \emph{agent names}, with
        access to input and output channels
        $\vin(a),\vout(a)\subseteq\channels$, respectively, such
        that $\vin(a)\cap \vin(a')=\emptyset$ for $a\neq a'$,
    \item
        $s_0$ is a finite sequence of output
        events, the \emph{initial output sequence}, for
        initializing parties, 
    \item
        for every $a\in A$,
        $P(a)\subseteq\Proc(\vin(a),\vout(a))$ is the set of
        \emph{programs} or \emph{processes of $a$}; this set is assumed to
        be closed under $\simeq$. 
    \end{ienum}
\end{definition}

For example, if $a$ is an honest
voter, then $P(a)$ would typically contain a program for
each way $a$ could vote, possibly including
abstention of voting. We note that the set $A$ typically
contains the coercer and coerced parties, i.e., these
entities are part of the protocol specification.

If $A = \{a_1,\dots,a_n\}$ and $p_{i} \in P(a_i)$, then
$(p_{1}\parallel \dots\parallel p_{n})$ is an
\emph{instance of $S$}, where the $p_1,\dots,p_n$ are
non-conflicting. A \emph{run of $S$} is a fair run of the
process $p_{1}\parallel \dots\parallel p_{n}$ initiated by
$s_0$, where $p_{1}\parallel \dots\parallel p_{n}$ is some
instance of $S$.

For a protocol $S = (A, \vin, \vout, s_0, P)$ and $a\in A$, a
channel $c$ is said to be \emph{private channel of $a$}, if $c
\in \vin(a)\cap\vout(a)$ and $c \notin \vin(a')\cup\vout(a')$
for all $a'\neq a$.

Now, let $S = (A, \vin, \vout, s_0, P)$ be a protocol with $A =
\{\voter,\coercer,\env\}$. Typically, $\env$ subsumes all honest
principals and processes in $P(\env)$ are of the form
$p_1\parallel \cdots \parallel p_n$, where $p_i$ are programs of
honest voters and authorities. Dishonest voters and authorities
are subsumed by the coercer $\coercer$ and coerced voters by
$\voter$.  For such a protocol we can define a coercion system,
as follows.

\begin{definition}
    Let $S = (A, \vin, \vout, s_0, P)$ be a protocol with $A =
    \{\voter,\coercer,\env\}$.
    \emph{The coercion system induced by $S$} is
    $(R,V,C,E,r,\sim)$, where
    \begin{ienum}
    \item
        $V = P(\voter)$, 
        $C = P(\coercer)$, and
        $E = P(\env)$, 
    \item
        $R$ is a set of tuples of the form
        $(v,c,e,\pi)$, with non-conflicting $v\in V$, $c\in C$, $e\in E$ and
        $\pi$ is a run induced by $(v \comp c \comp e)$.
    \item
        for every $v\in V$,
        $c\in C$, $e \in E$, $r(v,c,e) = \{(\hat v,\hat c,\hat e,\pi) \mid
        \hat v\simeq v$, $\hat c\simeq c$, $\hat e \simeq e$, and $\pi$ is a
        run of $S$ induced by $(\hat v\parallel \hat c\parallel \hat e)\}$ is
        the set of runs of the process formed by $v$, $c$, and $e$, closed under renaming of
        nonces, 
    \item
        for all $(v,c,e,\pi),(v',c',e',\pi')\in R$, we have
        $(v,c,e,\pi) \sim (v',c',e',\pi')$ iff $c = c'$ and $\pi
        \equiv_c \pi'$. Hence, the relation
        $\sim$ models the view of the coercer $c$ on runs of $S$.
    \end{ienum}
\end{definition}

    


    \section{General Properties}\label{sec:generalproperties}

In this section, we state general properties of coercion
systems induced by protocols, as introduced in the previous
section.  On the one hand, these properties facilitate
proofs of coercion resistance of voting protocols. On the
other hand, they demonstrate the adequacy of our model. In
Section~\ref{sec:dummytheorem}, we show that, under
reasonable assumptions, to prove coercion resistance it is
not necessary to consider all coercion strategies, i.e.,
all programs $v\in V$, but rather suffices to consider a
single coercion strategy, the dummy strategy. In
Section~\ref{sec:receiptfree}, we briefly discuss the
notion of receipt freeness and show that it is implied by
our notion of coercion resistance. We also show, in
Section~\ref{sec:multivotercoercion}, that multi-voter
coercion resistance, where multiple voters are coerced, is
implied by a slight extension of single-voter coercion
resistance, where only one voter is coerced.  Except for
the second statement, the other statements have not been
proven in other works on the symbolic analysis of voting
protocols.


\subsection{Dummy Theorem}\label{sec:dummytheorem}

The theorem that we want to prove, requires \emph{normal}
protocols. In these protocols the coerced voter and the
coercer can freely communicate (there are input and output
channels in both directions) and the set of programs of
both entities contains all processes, with appropriate
input and output channels.  For general coercion
resistance, protocols are typically defined in this way.


We define the \emph{dummy coercion strategy} $v_0$ to be
the process which simply forwards to the coercer all the
messages it receives from the environment and, conversely,
forwards to the environment all the messages it receives from
the coercer.

Now, we call a coercion system for a protocol \emph{dummy coercion
resistant} if it is coercion resistant in case a counter strategy is
demanded only for the dummy coercion strategy. 

To state our dummy theorem, we need to define a relation
$\peq$ on runs which defines a certain view of the environment.
Let $S=(A,\vin,\vout,s_0,P)$ be a protocol with $A =
\{\voter,\coercer,\env\}$ and let $\pi$ be a run of $P$.
Then, by $\envview(\pi)$ we denote the subsequence of $\pi$
which only contains input and outputs events for
channels of $\env$, i.e., events of the form $(c:m)$ and
$(\bar c,m)$ with $c\in \vin(\env)\cup\vout(\env)$.  Now,
for runs $\pi$ and $\pi'$ we write $\pi \peq \pi'$ iff
$\envview(\pi) = \envview(\pi')$. We extend this relation to the
set of runs of the coercion system of $S$: $(v,c,e,\pi)
\peq (v',c',e',\pi')$ iff $e=e'$ and $\pi\peq\pi'$.  We say
that a set $H$ of runs is \emph{closed under $\peq$},
if $\pi\in H$ and $\pi\peq\pi'$ implies
$\pi'\in H$. 

In the following theorem, we assume that $\alpha$ and
$\gamma$ are closed under $\peq$. As $\alpha$ and $\gamma$
are typically defined based on the view of the environment,
the assumption is satisfied in most applications, including
the protocols that we analyzed.
   
\begin{theorem}\label{th:dummy}
  Let $S=(V,C,E,r,\sim)$ be a coercion system for a normal
  protocol and $\alpha,\gamma$ be sets of runs of $S$
  closed under $\peq$. Then dummy coercion resistance
  implies (full) coercion resistance.
\end{theorem}
\begin{proof}[Proof sketch (see
  Appendix~\ref{sec:app-dummytheorem} for the full proof)]
  Assume that $v'_0$ is the counter strategy for the dummy
  strategy $v_0$. Let $v\in V$ be any coercion strategy.
  Then we show that the parallel composition of $v'_0$ and
  $v$, i.e., the process $v'_0\parallel v$, with a proper
  renaming of channels, is a counter strategy for $v$.
\end{proof}

We note that theorems of a similar flavor as the one above are also considered
in cryptographic, simulation-based settings (see, e.g.,
\cite{Canetti-eprint-067-2001,KuestersDattaMitchellRamanathan-JC-2008,Kuesters-CSFW-2006}).

\subsection{Receipt Freeness}\label{sec:receiptfree}

We define receipt freeness similarly to coercion
resistance, but with the assumption that the coercer cannot
send any messages directly to the coerced voter. Hence,
only the coerced voter can send messages to the coercer.
These messages can be considered as receipts.  This
intuition is shared with many other works.  One could
further weaken the following definition by fixing a certain
class of coercion strategies, where, for example, the
coerced voter basically follows the protocol but provides
the coercer with all the information obtained during the
run of the protocol.

    \begin{definition}\label{def:rf}
        A coercion system \mbox{$S=(V,C,E,r,\sim)$} is \emph{receipt-free} in
        $\alpha$ w.r.t.\ $\gamma$, if the system $S'=(V,C',E,r,\sim)$,
        where $C'$ consists of all the programs in $C$ which do not
        directly send messages to the coerced voter, is
        coercion-resistant in $\alpha$ w.r.t.\ $\gamma$.
    \end{definition}
%


    Alternatively to restricting the coercer, one could
    require the coerced voter not to accept messages from
    the coercer. As an immediate consequence of the above
    definition, we obtain the following theorem. 

    \begin{theorem}\label{the:coercionimpliesreceiptfree}
      If a coercion system is coercion-resistant, then it
      is receipt-free.
    \end{theorem}

\subsection{Multi-voter Coercion Resistance}\label{sec:multivotercoercion}

In this section, we show that multi-voter coercion
resistance is implied by a slight extension of single-voter
coercion resistance.  The main idea is that in case of
multiple coerced voters, all coerced voters, except for
one, can be considered to be dishonest, and hence, their
behavior can be subsumed by the coercer, leaving the case
of a single coerced voter.

In what follows, let $S = (A, \vin, \vout, s_0, P)$ be a
protocol with $A=\{\voter, \coercer, \env\}$.  According to
the definition of multi-voter coercion resistance (see
Section~\ref{sec:definingcoercionresistance}), we assume
that the programs of $\voter$ are processes of the form
$(p_1\comp\cdots \comp p_n)$, where $p_i$ represents a
process of the coerced voter $\voter_i$, with its own set
$I_i$ and $O_i$ of input and output channels, respectively.
We have that $\vin(\voter) = I_1\cup\dots\cup I_n$ and
$\vout(\voter) = O_1 \cup\dots\cup O_n$.

Given $S$, we define for every coerced voter $\voter_i$ a
new protocol $S_i$, where $\voter_i$ is the only coerced
voter and every other coerced voter is considered to be
dishonest, and hence, subsumed by the coercer. The
environment $\env$ in $S_i$ is the same as in $S$. 

    We let $T$ denote the coercion system for $S$ and $T_1,\dots,T_n$ the
    coercion systems for $S_1,\dots,S_n$,
    respectively.

    Now, we slightly extend the notion of
    (single-voter) coercion resistance, as mentioned
    before. An explanation follows the definition. 

    \begin{definition}\label{def:crex}
        A system $S=(R,V,C,E,r,\sim)$ is \emph{coercion resistant for
        $(\alpha_0,\dots,\alpha_n)$ w.r.t.\ $\gamma$}, where
        $\alpha_0,\dots,\alpha_n,\gamma\subseteq R$, if for
        each $v\in V$ there exists $v'\in V$ such that the
        following conditions are satisfied. 
        \begin{ienum}
        \item 
            For every $k\in\{1,\dots,n\}$, $c\in C$, $e\in E$, and $\rho \in
            r(v,c,e)\cap\alpha_k$, there exists $e'\in E$ and $\rho'
            \in r(v',c,e') \cap \alpha_{k-1}$ such that $\rho \sim \rho'$.
        \item
            For every $k\in\{1,\dots,n\}$, $c\in C$, $e\in E$, and $\rho \in
            r(v',c,e)\cap\alpha_k$, there exists $e'\in E$ and $\rho'
            \in r(v,c,e')\cap\alpha_{k-1}$ such that $\rho \sim \rho'$.
          \item For every $c\in C$ and $e\in E$, we have that $r(v',c,e) \subseteq \gamma$.
        \end{ienum}
    \end{definition}
    First note that condition (iii) of the definition is
    the same as the corresponding condition in
    Definition~\ref{def:cr}. Also, for $n=1$ and
    $(\alpha_0,\alpha_1)=(R,\alpha)$ the rest of the
    conditions coincide with Definition~\ref{def:cr} as
    well. A property $\alpha_i$ contains, for example, all
    runs in which there are at least $i$ votes for all
    candidates by honest voters. Now, when going from a run
    where the coerced voter carries out $v$ to a run where
    he/she carries out $v'$, then in the latter runs honest
    voters might have to vote in different ways in order to
    balance the behavior of $v'$. The above definition
    requires that in the run with $v'$ still $\alpha_{i-1}$
    is satisfied, and hence, in the example, there are
    still at least $i-1$ votes for all candidates by honest
    voters.

    We obtain the following theorem, which says that to
    prove multi-voter coercion resistance, it suffices to
    show single-voter coercion resistance in the sense of
    Definition~\ref{def:crex}. Despite the quantification
    over $i$ in the following theorem, it typically
    suffices to prove (single-voter) coercion resistance
    for one $T_i$, due to symmetry.

    \begin{theorem}\label{th:multiple}
      Let $S, S_1,\dots,S_n$ and $T, T_1,\dots,T_n$ be
      defined as above. Let $\alpha_0, \dots, \alpha_n$ and
      $\gamma_1, \dots, \gamma_n$ be properties of $T$,
      i.e., sets of runs of $T$.  If, for each
      $i\in\{1,\dots,n\}$, we have that $T_i$ is
      coercion-resistant for $(\alpha_0,\dots,\alpha_n)$
      w.r.t.\ $\gamma_i$, then $T$ is multi-voter coercion resistant in
      $\alpha_n$ w.r.t.\ $\gamma_1\cap\dots\cap\gamma_n$.
    \end{theorem}
    The proof of this theorem is postponed to
    Appendix~\ref{sec:proofmultiplecoercion}.  As mentioned
    before, the proof of this theorem relies on the fact
    that coerced voters, except for one, can be considered
    to be dishonest voters, and hence, can be subsumed by
    the coercer. Our analysis on the protocol by Okamoto
    \cite{Okamoto-SPW-1997} show that if dishonest voters
    are not considered, then single-voter coercion does in
    fact not imply multi-voter coercion: One can show that
    the Okamoto protocol is coercion resistant in the case
    of a single coerced voter without any dishonest voters.
    But the protocol is not coercion resistant with two
    coerced voters and still no dishonest voters (see
    Appendix~\ref{sec:okamoto}).



    \section{Civitas }\label{sec:civitas}

In this section, we briefly recall the Civitas system
\cite{ClarksonChongMyers-SP-2008}, discuss how this system
is modeled in our framework, and present positive and
negative results of our analysis of Civitas, i.e., we state
conditions under which Civitas does not guarantee coercion
resistance and conditions under which coercion resistance
is achieved. This is the first rigorous analysis of Civitas
and our analysis brings out subtleties that have not been
observed before. A detailed treatment can be found in
Appendix~\ref{sec:app-civitas}.

\subsection{Protocol Description}

We now briefly describe the Civitas system. A more detailed
specification of this system in our framework is provided
in the appendix. We start with a short description of the
various cryptographic primitives employed in Civitas.

\smallskip

\noindent \emph{Cryptographic primitives.}  Civitas uses,
among others, encryption schemes that allow for homomorphic
encryption, random reencryption, and/or distributed
decryption. In an encryption scheme with distributed
decryption, a public key is generated by multiple parties.
This public key can be used for encryption as usual.
However, the participation of all parties involved in
generating the public key is necessary to decrypt a message
encrypted under the public key.  Civitas also uses a
\emph{distributed plaintext equivalence test} (PET), where
multiple parties participate in determining whether two
different ciphertexts contain the same plaintext. Finally,
Civitas employs a number of zero-knowledge proofs and a mix
network. 

\smallskip

\noindent \emph{Protocol participants.} The Civitas system
assumes the following protocol participants: the supervisor
$\auts$, voters
$\voter_0,\dots,\voter_m$, the bulletin board $\autb$
(which is a kind of write-only, publicly accessible
memory), registration tellers $\autr_0,\dots,\autr_k$,
ballot boxes $\autbox_0, \dots,\autbox_k$, and tabulation
tellers $\autt_0,\dots,\autt_k$.  As in
\cite{ClarksonChongMyers-SP-2008}, we make the following
assumptions: $\auts$, $\autb$, $\autr_0$, $\autbox_0$, and $\autt_0$
are honest, the remaining voting authorities may be
dishonest.  An arbitrary number of voters are dishonest,
they are subsumed by the coercer. The channel between the
coerced voter and the honest registration teller is
untappable. Channels from voters to the ballot boxes are
anonymous, but not untappable (the coercer can see whether
ballots are sent to a ballot box).

For now, we consider one coerced voter, say $\voter_0$. We
note that in \cite{ClarksonChongMyers-SP-2008}, it is
assumed that $\voter_0$ knows which one of the registration
tellers is honest. It is in fact easy to see that Civitas
is not coercion resistant otherwise. We discuss the case of
multi-voter coercion at the end of this section.

\smallskip

\noindent \emph{Phases of the protocol.} The protocol has three
phases: the setup, voting, and tabulation phase. 

In the \emph{setup phase} the following steps are
performed.  The tabulation tellers collectively generate a
public key $\keyt$ and post it on the bulletin board;
messages encrypted under $\keyt$ are decrypted in a
distributed manner by the tabulation tellers.  Next, each
registration teller $\autr_j$ randomly generates, for each
voter $\voter_i$, a \emph{private credential share} $s_{ij}$ and
posts the corresponding public share
$S_{ij}=\parenc{s_{ij}}{\keyt}{r_{ij}}$ on the bulletin
board, where $r_{ij}$ represents the random coins used in
the encryption of $s_{ij}$.  The \emph{public credential
  $S_i$ of $\voter_i$} is publicly computable as $S_i =
(S_{i0}\times\dots\times S_{ik})$.  Now, a voter $\voter_i$
\emph{registers} at \emph{each} $\autr_j$ to acquire
his/her private credential shares $s_{ij}$, which comes with a
designated verifier reencryption proof (DVRP) that $s_{ij}$
corresponds to the public share $S_{ij}$ posted on the
bulletin board (such a proof is built using the public key
of the voter; a voter, or any party who knows the
corresponding private key, is able to forge such a proof,
which is crucial for coercion resistance). The voter then
computes his/her private credential $s_i = s_{i1} \times \dots
\times s_{ik}$.

In the \emph{voting phase}, a voter $v_i$ posts his
\emph{ballot} $b_i$ on all the ballot boxes (it is enough,
if the ballot is published on only one such a box to be
taken into account in the tabulation phase). A ballot
consists of an encrypted vote $\parenc{v}{\keyt}{r}$, the
encrypted credential $\parenc{s_i}{\keyt}{r'}$, a
zero-knowledge proof showing that $v$ is a valid vote, and a
zero knowledge-proof showing that the submitter
simultaneously knows $s_i$ and $v_i$.

In the \emph{tabulation phase}, tabulation tellers
collectively tally the election by performing the following
steps: (1) They retrieve the ballots from ballot boxes and
the public credentials from the bulletin board. (2) They
check the proofs of the ballots, eliminating those ballots
with invalid proofs. (3) Using PETs, duplicate ballots,
i.e., ballots with the same encrypted credential, are
eliminated according to some fixed policy. (4) First the
ballots and then the credentials are mixed by each
tabulation teller, by applying a permutation and using
reencryption. (5) Ballots without valid credentials are
eliminated, again using PETs. (6) The votes of the
remaining ballots are decrypted in a distributed manner by
the tabulation tellers and published. In steps (3)-(6)
zero-knowledge proofs are posted to ensure that these steps
are performed correctly.

\subsection{Negative Results}\label{sec:civitiasnegative}  Clarkson et al.~\cite{ClarksonChongMyers-SP-2008} claim that under the
assumptions mentioned before, Civitas is coercion
resistant.  Just as in the protocol by Juels et
al.~\cite{JuelsCatalanoJakobsson-WPES-2005}, the idea
behind the counter strategy of the coerced voter is to
provide the coercer with a fake credential, which prevents
the coercer from voting. Clarkson et al.~briefly mention
that a voter might not be able to vote if a registration
teller refuses to provide a credential share to the voter
and propose to use an additional voting authority, which
attest the misbehavior of the registration teller. However,
in the course of trying to prove that Civitas is coercion
resistant, we found further problems that make clear that,
under the mentioned conditions, Civitas does not provide
coercion resistance, if the goal of the coerced voter is to
vote for a specific candidate, say $z$.

The first problem is the following. We may well assume that
all dishonest registration tellers provide credential
shares to all voters. But they might in addition inform the
coercer who has registered. Now, if the coercion strategy
dictates the coerced voter not to register, there is no way
that the coerced voter can register, as the coercer would
be informed. In particular, there is no counter strategy
that would allow the coerced voter to vote for $z$, as the
coerced voter cannot register in the first place, and
hence, does not know all credential shares required for
casting a valid ballot.

There is also another more subtle coercion strategy, which
instructs the coerced voter to reveal his/her private key
to the coercer before the registration phase. Now, a
dishonest registration teller collaborating with the
coercer, can use this private key to forge the DVRP. As a result, the coerced
voter cannot be sure to have obtained a valid credential
share. Hence, even if this voter obtained a credential
share from every registration teller, he/she might still
not be able to vote.

\subsection{Positive Results}

We found that Civitas is
coercion resistant in all of the following three settings:
\begin{enumerate}
\item All registration tellers are honest and the goal of
  the coerced voter is to successfully vote for the
  candidate of his/her choice.
\item The goal of the coerced voter is only to prevent the
  coercer from casting a valid ballot, where otherwise the
  assumptions about channels and honest and dishonest
  authorities are as in \cite{ClarksonChongMyers-SP-2008}
  and discussed above.
\item The goal of the coerced voter is to successfully vote
  for the candidate of his/her choice, but the coercion
  strategies are restricted in that they first dictate the
  coerced voter to register as prescribed by the protocol
  and only then follow some arbitrary coercion strategy.
  Otherwise, the assumptions are as in
  \cite{ClarksonChongMyers-SP-2008} and discussed before.
\end{enumerate}
The assumptions in the first setting appear to be too
strong, given that the main difference of Civitas compared
to the Juels et al.~protocol, on which Civitas is based,
was to replace a single trusted registration teller by a
group of possibly dishonest registration tellers. The
second setting does not provide the coerced voter with much
guarantees.  The last setting, which we refer to by
\emph{Civitas with restricted coercion strategies}, seems
to be the most interesting and certainly the most
challenging to prove. We will therefore concentrate on this
setting in the rest of the section. One can imagine that
the registration is performed long before the election and
that in this phase the coercer does not yet try to
influence the voter.

We note that in case of Civitas with restricted coercion
strategies, the coercer can still ask the voter to reveal
his/her private key, but only after the registration of the
voter. Hence, the voter can check whether he/she has
obtained a valid credential share. Also note that
registration tellers might be dishonest.

The main theorem of this section states that Civitas with
restricted coercion strategies is coercion resistant in
$\alpha$ w.r.t.~$\gamma_z$ for any candidate $z$, in the
sense of Definition~\ref{def:cr}. We now formulate $\alpha$
and $\gamma_z$.

We first introduce some terminology. We say that a ballot posted
by a voter is \emph{posted successfully}, if this ballot is
delivered to the honest ballot box before the voting phase ends.
A run $\rho$ is \emph{fair} w.r.t.~the coerced voter $\voter_0$,
if, in this run, (1) all the registration and tabulation tellers
follow the protocol, i.e.\ post all messages and correct
zero-knowledge proofs, as required, (2) $\voter_0$ obtains his
credentials before the voting phase ends, and (3) if $\voter_0$
posts a valid ballot before the voting phase ends, then this
ballot is posted successfully.  

The properties $\gamma_z$ and $\alpha$ defined next, will
be discussed below.

For every candidate (or valid vote) $z$, the goal
$\gamma_z$ of the coerced voter $\voter_0$ is defined to be
the set of all runs satisfying the following conditions: If
a run is fair w.r.t.~$\voter_0$, then the coerced voter
successfully votes for $z$.

The set $\alpha$ of runs contains all runs satisfying the following
conditions: (1) For each possible candidate (or valid vote), there is
at least one honest voter who successfully casts this vote. (2) There
is at least one honest voter who obtains his credential before
$\voter_0$ finishes registration and abstains from voting. (3) There
is at least one honest voter who obtains his credential, but posts
successfully a ballot with an \emph{invalid} credential. (4) There is
at least one honest voter who posts a ballot after $\voter_0$ finishes
registration.

Let us first discuss $\gamma_z$. By
Definition~\ref{def:cr}, (iii) $\gamma_z$ means that the
counter strategy of $\voter_0$ must be such that $\voter_0$
votes successfully for $z$ in every fair run. In runs that
are not fair w.r.t.~$\voter_0$ it is clear that the vote of
the coerced voter will not be counted, either because a
tabulation teller misbehaved in an observable way, making
the election invalid, or the ballot did not reach any
ballot box in time, and as a result is not decrypted and published
by a tabulation teller. The latter can happen if messages
on the network are delayed for too long, possibly caused by
the coercer. These are obvious reasons why a vote might not
be counted. Hence, $\gamma_z$ is a very strong goal.

Now, consider the conditions (1) to (4) for $\alpha$:
Condition (1) was already motivated in
Section~\ref{sec:definingcoercionresistance}. Condition (2)
is needed because if no honest voter abstains from
voting, the coercer could tell that the coerced voter does
not abstain from voting, even though he/she was supposed to
abstain, just by counting the published votes.  Moreover,
if $\voter_0$ completed registration before everybody else
(the coercer can even force this to happen when cooperating
with a dishonest registration teller), then if some ballot
is posted, the coercer knows that this must have been
$\voter_0$. (We assume that honest voters do not post
ballots without completing registration.) In this way, the
coercer could again force $\voter_0$ to abstain from
voting. Condition (3) is also necessary. If the coercer
posts a ballot with the fake credential provided by
$\voter_0$, and if all honest voters only post valid
credentials, then the coercer can tell that he/she was
fooled, and hence, the counter strategy of the coerced
voter fails. Finally, condition (4) is needed for similar
reasons as condition (2).

Conditions (1) and (4) arguably exclude runs that are
unlikely to happen anyway. However, this is debatable for
condition (3) (maybe also for (2)). There is no reason to
assume that an honest voter would use an invalid
credential, even if he/she has a valid one (such a voter
would have to deviate from the protocol). To avoid
condition (3), we suggest that Civitas contains some
authority which randomly casts some ballots with invalid
credentials. Similar ``noise'' can also help to avoid
condition (2).

\begin{theorem} \label{th:civitas} The coercion system
  induced by Civitas with restricted coercion strategies is
  coercion resistant in $\alpha$ w.r.t. $\gamma_z$, for any
  valid vote $z$.
\end{theorem}

The proof of this theorem is given in the appendix.
Let us note that the theorem
holds for any number of honest and dishonest voters and
authorities.  We also note that the proof of this theorem
does not depend on the policy used to remove duplicates. In
particular, it does not matter whether re-voting is allowed
or not.

\myparagraph{Multi-voter coercion.}  Theorem
\ref{th:civitas} can easily be generalized to multi-voter
coercion resistance. Suppose that a number $k$ of voters is
being coerced. Suppose that the goal of voter $\voter_k$ is
$\gamma_k$.  By Theorem~\ref{th:multiple}, to prove
multi-voter coercion resistance in $\alpha$ w.r.t.\ $\gamma
= (\gamma_1\cap \cdots\cap \gamma_n)$, it is enough to
prove (*): a system with only one coerced voter $v_i$ is
coercion resistant for $(\alpha_0,\dots,\alpha_n)$
w.r.t.~$\gamma_i$, with $\alpha_n=\alpha$.
    
We define $\alpha_k$ as the set of runs
where (1) for each possible vote, there are at least $k$
honest voters who successfully cast this vote, (2) there
are at least $k$ honest voters who obtain their
credentials, before any of the coerced voters finishes
registration, and abstain from voting, (3) there are at least
$k$ honest voters who obtain their credential, but post ballots
with invalid credentials, (4) there are at least $k$ honest
voters who post a ballot after the coerced voters finish
registration.

The proof of (*) is very similar to the one for
Theorem~\ref{th:civitas}. Hence, multi-voter coercion
resistance follows.


    \section{Lee et al.~Protocol}\label{sec:leeprotocol}

In this section, we analyze a protocol proposed by Lee et
al.~\cite{LeeBoydDawsonKimYangYoo-ICISC-2003} within our
framework.  We show that the protocol is not coercion
resistant in general, but propose an extension of the
protocol for which we can show coercion resistance. 

\subsection{Protocol Description} The Lee et al.~protocol
assumes that every voter owns a tamper-resistant device,
called a \emph{randomizer}. 
    
In the \emph{setup phase}, the tallying tellers
$\autt_1,\dots,\autt_k$ generate and publish their common
public key $\keyt$ for threshold decryption. 

In the \emph{voting phase}, a voter prepares his/her
ballot, containing a vote encrypted under $\keyt$, and
gives it to his/her randomizer which reencrypts the ballot
and signs it, and sends the result $s$ back to the voter
along with a designated verifier reencryption proof (DVRP)
(such a DVRP can be forged by anyone who knows the private
key of the voter).  This part of the communication is
assumed to be entirely private.  Then the voter checks the proof,
computes his/her own signature on $s$ and posts it on the bulletin
board.

In the \emph{tallying phase}, the following is done: (1) the
double signatures of voters and their randomizers on the
posted ballots are verified and invalid ballots are
eliminated, (2) the remaining ballots are shuffled and
reencrypted and the result is posted on the bulletin board,
(3) talliers jointly decrypt shuffled ballots and publish
the tally result. Correctness of all these steps is assured
by posting appropriate non-interactive zero-knowledge
proofs.

\subsection{Negative Results} Assuming that the goal of
the coerced voter is to vote for a particular candidate, it
is easy to see that this protocol is not coercion
resistant: There is a simple abstention attack where the
coercer disallows the coerced voter to put a ballot signed
by this voter on the bulletin board.  So, one can at most
hope to prove that if a ballot signed by the coerced voter
and his/her randomizer has been put on the bulletin board,
then the vote of the coerced voter is counted. However,
even this weaker form of coercion resistance cannot be
shown: A coercer could prepare a ballot with some invalid
vote which is unlikely to occur otherwise and then ask the
coerced voter to give this ballot to his/her randomizer,
sign the result and put it on the bulletin board. The
coercer can check whether his/her vote is decrypted, assuming a
dishonest tallying teller collaborating with the
coercer. (The Lee et al.~protocol is designed to deal with
dishonest tallying tellers.) Therefore, a counter strategy
is forced to use the ballot prepared by the coercer, and
hence, the goal of the coerced voter cannot be achieved.

\subsection{Positive Results} To prove coercion
resistance, one could assume that all tallying tellers are
honest, but this is not the point of the Lee et
al.~protocol. We instead propose a slight extension of the
protocol, where the randomizer expects in addition to the
ballot a zero-knowledge proof which shows that the vote in
the ballot is well-formed (just as in Civitas).  The
randomizer then checks the proof before replying. With this
extension of the protocol, we obtain coercion resistance
for a natural $\gamma_z$ and $\alpha$: $\gamma_z$ contains
all runs where the coerced voter successfully votes for
$z$, if some ballot signed by this voter and his/her
randomizer appears on the bulletin board (within the voting
phase) and all zero-knowledge proofs that have to be
provided by the authorities are valid. Note that this goal
does not exclude abstention attacks. For the same reason
explained above, these attacks are still possible in the
extended version of the Lee et al.~protocol. The set
$\alpha$ is simply the set of runs where for each possible
vote there is at least one honest voter who successfully
casts this vote.

\begin{theorem} \label{th:lee} The coercion system induced
  by the extended version of the Lee et al.~protocol is
  coercion resistant in $\alpha$ w.r.t.  $\gamma_z$, for
  any valid vote $z$.
\end{theorem}

The proof of this theorem is sketched in the appendix.


    \section{Related Work}\label{sec:relatedwork}

Coercion resistance in a symbolic model was first
formulated by Delaune et
al.~\cite{DelauneKremerRyan-CSFW-2006,DelauneKremerRyan-JCS-2008,DelauneKremerRyan-WOTE-20006}.
This work was then further developed by Backes et
al.~\cite{BackesHritcuMaffei-CSF-2008}. Both the work by
Delaune et al.~and Backes et al.~were motivated by the
desire to use ProVerif \cite{Blanchet-CSFW-2001}, a tool
for security protocol analysis, for the automatic analysis
of voting protocols. Due to the focus on automation, the
notions of coercion resistance studied in these works are
more restricted than the one considered here. For example,
the notion of coercion resistance introduced by Delaune et
al.~does not apply to Civitas or the protocol by Juels et
al.\cite{JuelsCatalanoJakobsson-WPES-2005}, as the class of
coercion strategies and counter strategies they consider
are too restricted.  To show coercion resistance of the Lee
et al.~protocol, Delaune et al.~study a variant of this
protocol which is different to the one studied here. One of
the abstention attacks that we point out still works for
their variant. However, this attack is out of the scope of
their notion of coercion resistance. Conversely, the notion
of coercion resistance by Backes et al.~is inspired by the
one of Juels et al., which in turn is especially tailored
to the specific protocol structure of the protocol by Juels
et al.~and the specific forms of coercion strategies.  In
order to facilitate automation, the protocol models that
Delaune et al.~and Backes et al.~consider are much coarser
than ours.  For example, the way votes are tallied is
simplified and mix networks and proofs of compliance are
not modeled.

There is also a more fundamental difference between the
work by Delaune et al.~and Backes et al.~on the one hand,
and our work on the other hand. The symbolic model by
Delaune et al.~and Backes et al.~is the applied pi calculus
\cite{AbadiFournet-POPL-2001}, with its notion of
observational equivalence for comparing systems/processes.
Observational equivalence is a bisimilarity relation which
demands that every step of one system is matched by a
similar step of the other system. In particular, in the
works by Delaune et al.~and Backes et al.~the two systems
in which the coerced voter runs the coercion strategy and
the counter strategy, respectively, are related using the
notion of observational equivalence.  This is fundamentally
different to the approach taken here: In our epistemic
approach, we relate \emph{traces} of systems and say that
for every trace of one system, there exists a trace of the
other system such that the coercer has the same view on
both traces. In the two traces, honest voters may vote in
different ways. By this, votes (including abstention) can
be balanced in case coerced voters vote in different ways
in the two systems and this balancing may be based on the
traces as a whole.  Conversely, observational equivalence,
with its strict stepwise correspondence between systems,
prohibits a simple balancing of votes. As a result, the
formulations of coercion resistance proposed by Delaune et
al.~and Backes et al.~are very complex and less intuitive.
In Delaune et al., the balancing problem is tackled by
restricting the set of coercers and coercion strategies. It
is assumed that the coercer's goal is to vote for a
particular party and that coercion strategies only slightly
deviate from the prescribed protocol.  Altogether this
leads to a rather weak notion of coercion resistance,
excluding, for example, abstention attacks and other
natural coercion strategies, e.g., those relevant for
Civitas. Backes et al.~introduce what they call an
extractor to solve the balancing problem, which makes the
definition of coercion resistance quite complex and hard to
understand.

In \cite{JonkerPieters-WOTE-2006,JonkerVink-ISC-2006},
Jonker et al.~also follow an epistemic approach to model
properties of voting protocols. However, they do not
consider coercion resistance, only receipt freeness.
Receipt freeness is modeled w.r.t.~a message that a voter
could use as a receipt. This is only a very rough
approximation of the intuition behind receipt freeness.
Also, Jonker et al.~do not model any cryptographic
operators. A more recent work on receipt freeness by Jonker
et al.~is \cite{JonkerMauwPan-ARES-2009}.

The work by Baskar et al.~\cite{BaskarRamanujamSuresh-TARK-2007}
focuses on the decidability of knowledge-based properties of voting
protocols. However, they only study a very simplistic notion of
receipt-freeness, which resembles privacy of
votes; coercion resistance is not considered.

As already mentioned in the introduction, there also exist
several cryptographic definitions of coercion resistance
and receipt freeness (see, e.g.,
\cite{Okamoto-SPW-1997,JuelsCatalanoJakobsson-WPES-2005,MoranNaor-CRYPTO-2006,Chevallier-MamesFouquePointchevalSternTraore-WOTE-2006,TeagueRamchenNaish-WOTE-2008}).
On the one hand, compared to the cryptographic definitions,
our symbolic approach abstracts from many cryptographic
details, including details of cryptographic primitives and
probabilistic aspects. This leads to weaker security
guarantees. On the other hand, the simplicity of the
symbolic approach in general, and our definition in
particular, facilitates the analysis of protocols and is
more amenable to automation, which, given the complexity of
voting protocols, is a crucial advantage.


    \section{Conclusion}\label{sec:conclusion}

In this paper, we presented a general, yet simple and
intuitive definition of coercion resistance of voting
protocols in an epistemic setting, which does not depend on
any specific, symbolic protocol or adversary model.  We
applied our definition to three different voting protocols,
two of which, namely Civitas and the protocol by Okamoto,
have not been rigorously analyzed before. For all three
protocols, we identified conditions under which these
protocols are coercion resistant or fail to be coercion
resistant. To obtain these results it was vital that our
definition of coercion resistance allows to specify various
degrees of coercion resistance in a way more fine-grained
than in previous proposals. Our analyzes brought out
several insights about the three protocols that have not
been observed before and that led us to propose
improvements of the protocols.

We believe that our definition of coercion resistance
provides a good basis for automated analysis of coercion
resistance, in particular since the definition can be
instantiated with different protocol and adversary models.
However, carrying out tool supported analysis was out of
the scope of the present work.


{\small
\bibliography{literature}
}
\bibliographystyle{plain}

\appendix

\section{General Properties}\label{sec:app-generalproperties}

\subsection{Proof of Theorem \ref{th:dummy}}\label{sec:app-dummytheorem}

Before we present the proof of Theorem~\ref{th:dummy}, we
define normal protocols precisely.

\begin{definition}\label{def:app-normalprotocol}
  A protocol $(A,\vin,\vout,s_0,P)$ with $A =
  \{\voter,\coercer,\env\}$ is \emph{normal}, if (i)
  $\voter$ and $\coercer$ are connected by some input and
  output channels (in both directions), (ii) both $\voter$
  and $\coercer$ have an unbounded number of private
  channels (see the paragraph after
  Definition~\ref{def:protocol}), (iii)
  $P(\voter)=\Proc(\vin(\voter),\vout(\voter))$ and
  $P(\coercer)=\Proc(\vin(\coercer),\vout(\coercer))$.
\end{definition}

\myparagraph{Proof of Theorem \ref{th:dummy}.}  We first
introduce some terminology and prove general lemmas about
processes for forwarding messages between channels.

Let $p =
(I,O,N,f)$ be an atomic process and $h$ be a \emph{channel
  renaming}, i.e.\ a injection from $\channels$ to
$\channels$. We define an atomic process $h(p)$ as $(I',
O', N, f')$ with $I'=h(I)$ and $O'=h(O)$, where, for each
$(I',N)$-view $U = (\lambda,W)$, we put $f'(U) = h(f(
h^{-1}(\lambda), W ))$.  We extend the domain of a channel
renaming to arbitrary (non necessarily atomic) processes in
a natural way.

    Now, for processes $P_1,P_2$ and a channel renaming $h$, we will
    write $P_1 \sqsubseteq_h P_2$, if for each run $\pi$ induced by
    $P_1$, the run $h(\pi)$ is induced by $P_2$.  The following lemma
    is easy to prove. 

    \begin{lemma}\label{lem:remap}
        For a process $P$ and a channel renaming $h$, we have 
        $P \sqsubseteq_h h(P)$.
    \end{lemma}

    For $c,d\in\channels$, we denote by $\forw cd$ the process  which
    simply forwards on channel $d$ every message received on $c$. By
    $\forw cd\forw{c'}{d'}$ we will denote $(\forw cd \; \parallel \;
    \forw{c'}{d'})$.  For a process $P[c_0,c_1]$ which uses channels
    $c_0$ and $c_1$, we will write $P[a_0,a_1]$ for the process which
    uses $a_i$ instead of $c_i$ and otherwise behaves like
    $P[c_0,c_1]$ (i.e.\ $P[a_0,a_1] = h(P[c_0,c_1])$ for $h =
    \{c_0\mapsto a_0, c_1\mapsto a_1\}$).  Sometimes we will write
    $P[\vec c]$ instead of $P[c_0,c_1]$.


    Let $P$ be a subprocess of some process $P'$.  We define an
    equivalence relation $\peq_P$ on runs induced by $P'$ as follows:
    $\pi \peq_{P} \pi'$ iff $\pi_{|D} = \pi'_{|D}$,
    where $D$ is the set of elements of the form $c$ and $\bar c$, for
    $c\in \inp{P}\cup\outp{P}$. 
    Note that, if $\pi$ and $\pi'$ are 
    runs of some protocol $S=(A,\vin,\vout,s_0,P)$ with
    $A = \{\voter,\coercer,\env\}$ and $P \in P(\env)$, 
    then $\pi \peq \pi'$ iff $\pi \peq_P \pi'$.
    
    Let $P$ be a process. A channel $c$ is called an \emph{input
    channel of $P$}, if $c\in\inp{P}$ and $c\notin\outp{P}$.  A
    channel $c$ is called an \emph{output channel of $P$}, if
    $c\in\outp{P}$ and $c\notin\inp{P}$.

    \begin{lemma}\label{lem:forward}
        Let $P[c_0,c_1]$ be a process with some input channel $c_0$
        and some output channel $c_1$, let $P'$ be a process, and
        $x_0,x_1$ be channels not used neither by $P[c_0,c_1]$ nor
        $P'$. Let $P_1 = (P' \parallel P[c_0,c_1])$ and
        $P_2 = (P' \parallel \forw{c_0}{x_0}\forw{x_1}{c_1}
        \parallel P[x_0,x_1])$. 
        Then:
        \begin{enumerate}
        \item[(1)]
            For each run $\pi$ induced by $P_1$ there exists a run
            $\pi'$ induced by $P_2$ with $\pi\peq_{P'}\pi'$.
        \item[(2)]
            For each run $\pi$ induced by $P_2$ there exists a run
            $\pi'$ induced by $P_1$ with $\pi\equiv_{P'}\pi'$.
        \end{enumerate}
    \end{lemma}
    \begin{proof}
        Let $f$ be the process
        $\forw{c_0}{x_0}\forw{x_1}{c_1}$.  To prove (1), suppose that
        $\pi$ is a run induced by $P_1$.  We construct $\pi'$ in the
        following way.  Whenever it happens in $\pi$ that $(c_0:m)$ is
        delivered and, in consequence, a reply of $P[c_0,c_1]$ is
        sent, then two steps are performed in $\pi'$: first, $(c_0:m)$
        is delivered and, in consequence, the reply $(x_0:m)$ of $f$
        is sent; and second, the message $(x_0:m)$ sent in the first
        step is immediately delivered and, in consequence, the reply
        of $P[x_0,x_1]$ is sent.  Furthermore, whenever it happens in
        $\pi$ that $(c_1:m)$ is sent by $P[c_0,c_1]$, then two steps
        are performed in $\pi'$: the corresponding message $(x_1:m)$
        is sent by $P[x_0,x_1]$ and then this message is immediately
        delivered and, in consequence, $(c_1:m)$ is sent by $f$.  It
        is easy to show that $\pi'$ obtained in this way is a run
        induced by $P_2$ and $\pi' \peq_{P'} \pi$.

        To prove (2), suppose that $\pi$ is a run
        induced by $P_2$. We construct $\pi'$ in the following way.
        Whenever it happens in $\pi$ that $(x_1:m)$ is sent by
        $P[x_0,x_1]$, then, in $\pi'$, the corresponding message
        $(c_1:m)$ is sent by $P[c_0,c_1]$. When, in $\pi$, such a
        message $(x_1:m)$ is delivered and, in consequence, the reply
        $(c_1:m)$ of $f$ is sent, no corresponding step is performed
        in $\pi'$, so, in particular, $(c_1:m)$ is kept as a message
        to be delivered.  Furthermore, whenever it happens in $\pi$
        that a message $(c_0:m)$ is delivered and, in consequence, the
        reply $(x_0:m)$ of $f$ is sent, then no corresponding step is
        taken in $\pi'$, so, $(c_0:m)$ is kept as a message to be
        delivered.  When, in $\pi$, a message $(x_0:m)$ is delivered
        to $P[x_0,x_1]$, then, in $\pi'$, we can deliver $(c_0:m)$ to
        $P[c_0,c_1]$.  It is easy to show that $\pi'$ obtained in this
        way is a run induced by $P_1$ and $\pi' \equiv_{P'} \pi$.
        (Note, however, that one cannot prove $\pi'
        \peq_{P'} \pi$.) This completes the proof of
        Lemma~\ref{lem:forward}. 
    \end{proof}

    To make the proof of Theorem~\ref{th:dummy} simpler, we
    assume that $\vec a = (a_0,a_1)$ with
    $a_0\in\chout(\voter)\cap\chin(\coercer)$ and
    $a_1\in\chout(\coercer)\cap\chin(\voter)$, are the only
    channels shared by $\coercer$ and $\voter$ in a normal
    protocol. Similarly, we assume that $\vec d =
    (d_0,d_1)$ with $d_0\in\chout(\env)\cap\chin(\voter)$
    and $d_1\in\chout(\voter)\cap\chin(\env)$ are the only
    channels shared by $\env$ and $\voter$. We stress, that
    these assumptions make the proof simpler, but are by no
    mean crucial and can be easily dropped.

    For channels $\vec x = (x_0,x_1)$, let $v_0[\vec x]$ be
    $(\forw{d_0}{x_0}\forw{x_1}{d_1})$.  So, $v_0[\vec x]$  simply
    forwards on channel $x_0$ each message received on $d_0$ and
    forwards on $d_1$ each message on channel $x_1$.  Now, $v_0$ is
    just $v_0[\vec a]$. 

    \medskip
    To prove Theorem \ref{th:dummy}, suppose that $v_0$ is not a
    coercion strategy in $\alpha$ w.r.t.\ $\gamma$ and $v'_0$ is a
    counter-strategy for $v_0$.  Let $v$ be a strategy in $V$. We will
    construct a counter-strategy $v'$ for $v$.

    We will write $v_0'[\vec a]$ instead of $v_0'$, as channels $\vec
    a$ are used by $v_0'$. Similarly, we will write $v[\vec d,\vec
    a]$ and $c[\vec a]$, for any $c\in C$.  Let $\vec x
    = (x_0,x_1)$ be some private channels of $\voter$ not used in $v$
    nor $v'_0$.  Such channels exist due to
    Condition~(ii) of Definition~\ref{def:app-normalprotocol}.
    Similarly, for a given $c\in C$, let $\vec y = (y_0,y_1)$ be
    some internal channels of $\coercer$ not used in $c$.  
    
    We define $v'$ as $(v'_0[\vec x] \parallel v[\vec x,\vec a]).$  We will
    show that $v'$ is a counter-strategy for $v$.   Let $\sigma = \{
    \vec x \mapsto \vec a, \vec a\mapsto \vec y, \vec y \mapsto x \}$.   


    The following lemma holds true, because
    none of $\vec a, \vec x, \vec y$ is used by any $\hat e \in E$.

    \begin{lemma}\label{lem:eeq}
        Let $\rho = (\hat v, \hat c, \hat e, \pi)$ be a run of $S$.
        We have that $\pi \peq \sigma(\pi)$.
    \end{lemma}

        %
    \begin{lemma}\label{lem:mapsim}
        Let $\pi_1,\pi_2$ be runs induced by some $(\hat v \comp \hat c
        \comp \hat e)$ such that channels $\vec x$ do not occur in
        $\pi_1,\pi_2$. If  $\pi_1 \equiv_{\hat c} \pi_2$, then
        $\sigma^{-1}(\pi_1) \equiv_{\hat c} \sigma^{-1}(\pi_2)$.
    \end{lemma}
    \begin{proof}[Sketch of proof]
        The lemma follows from the observation, that, for each channel
        $z$ occurring in $\pi_1$ or $\pi_2$ (note that $z\neq x$), if
        $\sigma^{-1}(z) \in \inp{\hat c}$, then
        $z \in \inp{\hat c}$.
    \end{proof}

    Now we will show that Item (iii) of the definition of coercion
    resistance holds for $v'$, i.e.\ $r(v',c,e)\subseteq\gamma$,
    for all $c\in C$ and $e \in E$.  So, let $\rho_1  \in r(v',c,e)$,
    which means that 
    $$
        \rho_1 = ((\hat v'_0[\vec x] \parallel \hat v[\vec x,\vec a]),
        \hat c[\vec a], \hat e, \pi),
    $$
    for some $\hat v'_0 \simeq v_0$, $\hat v \simeq v$, $\hat c \simeq
    c$, $\hat e \simeq e$, and some $\pi$ induced by $(\hat v'_0[\vec
    x] \parallel \hat v[\vec x,\vec a] \comp \hat c[\vec a] \comp \hat
    e)$. So, by Lemma \ref{lem:remap}, $\sigma(\pi)$ is a run induced
    by $( \hat v'_0[\vec a] \comp \hat v[\vec a, \vec y] \comp \hat
    c[\vec y]\comp \hat e)$ and thus 
    $$
        \rho_2 =(\hat v'_0[\vec a], (\hat v[\vec a,\vec y] \parallel
        \hat c[\vec y]), \hat e, \sigma(\pi))
    $$
    is in $r(v_0'[\vec a], (v[\vec a,\vec y] \parallel c[\vec y]),
    e)$ (note that $v[\vec a,\vec y] \parallel c[\vec y]$ is in
    $P(\coercer)$, by condition (iii) of Definition~\ref{def:app-normalprotocol}).
    Because $v'_0[\vec a]$ is a counter-strategy for $v_0[\vec
    a]$, we have that $\rho_2 \in \gamma$.  By Lemma~\ref{lem:eeq},
    $\pi \peq \sigma(\pi)$, which implies $\rho_1 \peq
    \rho_2$. Because $\gamma$ is closed under $\peq$, we obtain
    $\rho_1 \in \gamma$.

    \medskip
    Finally, we will show that Item (ii) of the definition of coercion
    resistance holds for $v$ and $v'$, i.e.\
    for each $c$, $e$, and $\rho\in r(v',c,e)\cap\alpha$, there exist $e'\in E$
    and $\rho' \in r(v,c,e)$ such that $\rho\sim\rho'$.
    For Item (i) one can proceed
    similarly. This completes the proof of the theorem.

    So, let $\rho_1 \in r(v',c,e)\cap\alpha$. We proceed, as above,
    and so, $\rho_1$ is like above and, for $\rho_2$ defined as above,
    $\rho_1 \peq \rho_2$ holds.  Because $\alpha$ is closed under
    $\peq$, we have $\rho_2 \in \alpha$.  As $v'_0[\vec a]$ is a
    counter-strategy for $v_0[\vec a]$, there exists $e'\in E$ and a
    run $\rho_2'\in r(v_0[\vec a], (v[\vec a,\vec y] \parallel c[\vec
    y]), e')$ with $\rho_2 \sim \rho_2'$.  This means that 
    $$
        \rho'_2 = (v_0[\vec a], (\hat v[\vec a,\vec y] \parallel \hat
        c[y]), \hat e', \pi'),
    $$ 
    for some $\hat e' \simeq e'$, and some $\pi'$ induced by
    $(v_0[\vec a] \comp \hat v[\vec a,\vec y] \parallel \hat c[y]
    \comp \hat e')$ such that $\pi' \equiv_{(\hat v[\vec a,\vec y]
    \comp \hat c[y])} \sigma(\pi)$.  By Lemma~\ref{lem:remap},
    $\sigma^{-1}(\pi')$ is a run induced by $(v_0[\vec x] \parallel
    \hat v[\vec x,\vec a] \comp \hat c[\vec a] \comp \hat e')$ and
    thus 
    $$
        \rho'_1 = ( (v_0[\vec x] \parallel \hat v[\vec x,\vec a]),
        \hat c[\vec a], \hat e', \sigma^{-1}(\pi'))
    $$
    is in $r( (v_0[\vec x] \parallel v[\vec x,\vec a]),  c[\vec a],
    e')$ (note that $v_0[\vec x] \parallel v[\vec x,\vec a]$ is in
    $P(\voter)$, because of condition (iii) of Definition
    \ref{def:app-normalprotocol}).  Since $\pi' \equiv_{(\hat v[\vec a,\vec y]
    \comp \hat c[y])} \sigma(\pi)$ and $\inp{\hat c[\vec a]} \subseteq
    \inp{\hat v[\vec a,\vec y] \comp \hat c[y]}$, we have $\pi'
    \equiv_{\hat c[\vec a]} \sigma(\pi)$. So by Lemma
    \ref{lem:mapsim}, $\sigma^{-1}(\pi') \equiv_{ \hat c[\vec a]}
    \pi$.  Now, by Lemma~\ref{lem:forward}, there exists a run $\pi''$
    induced by $(\hat v \parallel \hat c[\vec a] \parallel \hat e')$
    such that $\pi'' \equiv_{(\hat c[\vec a]\comp\hat e')}
    \sigma^{-1}(\pi')$ with implies $\pi'' \equiv_{\hat c[\vec a]}
    \sigma^{-1}(\pi')$.  Hence, $\pi'' \equiv_{\hat c[\vec a]} \pi$,
    and so, finally, we obtain a run $(\hat v, \hat c, \hat e', \pi''
    ) \sim \rho_1$ in $r(v,c,e')$.



\subsection{Proof of Theorem~\ref{th:multiple}}\label{sec:proofmultiplecoercion}

Before we prove the theorem, we state some definitions only
sketched or omitted in
Section~\ref{sec:multivotercoercion}.

Let $S$ be a protocol as in
Section~\ref{sec:multivotercoercion}. We define $S_i= (A,
\vin_i,\vout_i, s_0, P_i)$, where $\voter$ now represents
voter $\voter_i$ only, $\env$ is unchanged, and $\coercer$
gets direct access to the channels of the coerced voters
$\voter_1,\dots, \voter_{i-1},
\voter_{i+1},\dots,\voter_n$, i.e., $\vin_i(\voter) = I_i$,
$\vout_i(\voter) = O_i$, $\vin_i(\coercer) = \vin(\coercer)
\cup \bigcup_{i\in W} I_i$, and $\vout_i(\coercer) =
\vout(\coercer) \cup \bigcup_{i\in W} O_i$, where
$W=\{1,\ldots,n\}\setminus \{i\}$.  Moreover, $P_i(\env) =
P(\env)$, $P_i(\voter)=
\Proc(\vin_i(\voter),\vout_i(\voter))$, and $P_i(\coercer)=
\Proc(\vin_i(\coercer),\vout_i(\coercer))$.

For the proof of Theorem~\ref{th:multiple}, we define a
mapping from runs $\rho$ of $T$ to runs
$\rho^{(i)}$ of $T_i$ and from properties $\beta$ of $T$ to
properties $\beta^{(i)}$ of $T_i$: Recall that each $v\in
P(\voter)$ is of the form $(v_1\comp\cdots\comp v_n)$ with
$v_i \in \Proc(I_i,O_i)$.  For a run $\rho =
((v_1\comp\cdots\comp v_n),c,e,\pi)$, we define
$\rho^{(i)}$ as $(v_i, (v_1\comp \dots\comp v_{i-1}\comp
v_{i+1}\comp \dots\comp v_n\comp c), e,\pi)$.  For a property
$\beta$ of $T$, we define $\beta^{(i)}$ to be $\{
\rho^{(i)} : \rho\in\beta \}$. When it is clear from the
context, we will write $\beta$ instead of $\beta^{(i)}$,
treating $\beta$ as a property of $T_i$.

We can now turn to the proof of Theorem~\ref{th:multiple}. We define a
function $f$ which maps a coercion strategy $v_i$ of the $i$-th voter to a
counter strategy $v_i' = f(v_i)$, by defining $v'_i$ as some
(arbitrarily chosen) counter strategy for $v_i$ in $T_i$ (such a
counter strategy exists, since $T_i$ is coercion resistant).

Now, for any $v\in P(\voter)$ which, as we know, must be of
the form $(v_1 \comp \dots \comp v_n)$ with $v_i \in
P_i(\voter)$, and for $v' = (v'_1 \comp \dots \comp v'_n)$,
where $v'_i = f(v_i)$, we will show that $T$, along with
$v$ and $v'$, meets the conditions of the definition of
multi-voter coercion resistance.
    
    First, let us show that condition (iii) holds. Let $c\in
    P(\coercer)$, $e\in P(\env)$ and $\rho \in r(v',c,e)$.  So, $\rho$
    is of the form $( (\hat v'_1 \comp\dots \comp \hat v'_n), \hat c,
    \hat e, \pi)$. For each $i\in\{1,\dots,n\}$ we have that
    $\rho^{(i)} \in r_i(v'_i, c_i, e)$, where $c_i=(v_1\comp
    \dots\comp v_{i-1} \comp v_{i+1} \comp\dots \comp v_n \comp c)$.
    Thus, $\rho^{(i)} \in \gamma_i$ and so $\rho\in\gamma_i$. Hence,
    $\rho \in \gamma_1 \cap\dots\cap \gamma_n$.

    Now, let us show that condition (i) holds. The proof for condition
    (ii) is very similar.  Let $\rho\in r(v,c,e) \cap \alpha_n$, for
    some $c$ and $e$.  Let $u_k$ denote $(v'_1 \comp\dots \comp v'_k
    \comp v_{k+1} \comp\dots \comp v_n)$.  Note that $u_0 = v$ and
    $u_n = v'$.  We will show, by induction, that for each
    $k\in\{0,\dots,n\}$ there exists $e_k$ and $\rho_k$ such that
    $\rho_k \in r(u_k, c, e_k) \cap \alpha_{n-k}$ and $\rho_k \sim
    \rho$.  Note that, for $k=0$, we can simply take $e_0=e$ and
    $\rho_0 = \rho$.  So, let us assume that the above holds for
    $k-1$. We will show that it also holds for $k$. 
    So, we have some $e_{k-1}$ and $\rho_{k-1}\sim\rho$ such that
    $\rho_{k-1} \in r(u_{k-1}, c, e_{k-1}) \cap \alpha_{(n-k+1)}$.  It
    follows that $\rho_{k-1}^{(k)} \in \alpha_{(n-k+1)}$ and
    $\rho_{k-1}^{(k)} \in r_i(v_{k},c^*,e_{k-1})$, where $c^* = (v'_1
    \comp\dots\comp v'_{k-1} \comp v_{k+1}\comp\dots \comp v_n \comp c)$.  By
    coercion resistance of $T_k$, there exists $e_k$ and $\rho'_k \in
    r(v'_k, c^*, e_k) \cap \alpha_{n-k}$ such that $\rho'_k \sim_k
    \rho_{k-1}^{(k)}$. Let $\rho_k$ be such that $\rho_k^{(k)} =
    \rho'_k$.  Hence, $\rho_k \sim \rho_{k-1}$ (as the coercer can see
    more in $T_k$ than in $T$). By transitivity of $\sim$, we have
    $\rho_k\sim\rho$.  We also have that $\rho_k \in r(u_k,c,e_k)$ and
    $\rho_k \in \alpha_{n-k}$.


\section{Civitas}\label{sec:app-civitas}

In this section we provide a detailed modeling of Civitas in our
framework and present the proof of coercion resistance of this system.

\subsection{Cryptographic Primitives}

    We use a term of the form $\an{m,m'}$ to represent a pair of messages $m$ and
    $m'$; with $\first p$ and $\second p$ yielding, respectively, the
    first and the second component of a pair $p$. A term $\sig km$
    represents the signature on a message $m$ under a (private) key
    $k$. Such a signature can be verified using $\pub(k)$, the public
    key corresponding to $k$. We also assume that such a signature
    reveals $m$. 
    
    We use the following terms to represent randomized encryption with
    reencryption and homomorphic property: $\parenc mkr$ represents a
    term $m$ encrypted under a (public) key $k$ using a randomness
    $r$; $\dec ck$ represents a decryption of a ciphertext $c$ with a
    key $k$ ($k$ is intended to be a private key corresponding to the
    public key under which $c$ is encrypted); $\reenc(c,k,r)$
    represents a reencryption of a ciphertext $c$ under a (public) key
    $k$ with randomness $r$ (we have $\reenc({\parenc mkr},k,r') =
    \parenc mk{r+r'}$). We also use symbols $+$ and $\times$, equipped
    with the appropriate equational theory, to
    express the \emph{homomorphic} property of the encryption:
    $\parenc{m_1}k{r_1} \times \parenc{m_2}k{r_2} = \parenc{m_1\times
    m_2}k{r_1+r_2}$.
    
    \emph{Distributed decryption} is modelled as follows. Suppose that
    $x_1,\dots,x_n$ are \emph{private key shares} of some agents
    $a_1,\dots,a_n$. Then, $\pub(x_1),\dots,\pub(x_n)$ are the
    corresponding \emph{public key shares} (which are intended to be
    published). The \emph{distributed public key} of $a_1,\dots,a_n$
    is now $K = \pub(x_1)\times \cdots \times\pub(x_n)$. To decrypt a
    ciphertext $c = \parenc mKr$, that is a message $m$ encrypted
    under this key, the cooperation of all $a_1,\dots,a_n$ is
    necessary: each $a_i$ posts his \emph{public decryption
    share} $p_i = \dpart(c,x_i)$.  Now, the result of decryption (that
    is the message $m$) can be computed from these shares: $m =
    \distdec(p_1,\dots,p_n)$.

    In a very similar way me model \emph{distributed 
    plaintext equivalence test} (PET), which can be used to
    determine, whether, for two ciphertext $c$ and $c'$, the plaintext
    of $c$ and $c'$ are the same, without revealing anything more about
    these plaintexts (in particular, without decrypting $c$ and $c'$).
    Suppose, again that
    $x_1,\dots,x_n$ are private key shares of 
    $a_1,\dots,a_n$ and $K = \pub(x_1)\times \cdots \times\pub(x_n)$
    is their distributed public key. To perform a PET on ciphertexts
    $c = \parenc mKr$ and
    $c' = \parenc {m'}K{r'}$ (that is to check whether $m$ and $m'$
    are the same),
    each $a_i$ posts his \emph{public PET share}
    $p_i = \petpart(c,c',x_i)$.  Now, the result of the PET
    can be computed from these shares: 
    $\distpet(p_1,\dots,p_n) = \True$ iff the $m=m'$.

    \begin{figure*}[t]\centering
    \parbox[b]{.55\textwidth}{
    \begin{align*}
        \checksig{\sig{m}{k}}{\pub(k)} &= \True \\[1ex]
        \unsig{\sig{m}{k}} &= m \\[1ex]
        \dec{\parenc x{\pub(k)}r}{k} &= x \\[1ex]
        \reenc(\parenc xkr,k,r') &= \parenc{x}{k}{r+r'} \\[1ex]
        \reenc(\reenc(x,k,r),k,r') &= \reenc(x,k,r+r') \\[1ex]
        \parenc{m_1}{k}{r_1} \times \parenc{m_2}{k}{r_2}  &= \parenc{m_1 \times m_2}{k}{r_1 + r_2} \\[1ex]
        \distdec(p_1, \dots, p_k) &= m \quad
            \parbox[t]{14em}{\small where $p_i = \dpart( \parenc{m}{Y}{r},
            x_i)$\\ with $Y = (\pub(x_1)\times\dots\times\pub(x_k))$}\\[1ex]
        \distpet(p_1, \dots, p_k) &= \True \quad
            \parbox[t]{15.5em}{\small where $p_i = \petpart(
            \parenc{m}{Y}{r}, \parenc{m}{Y}{r'}, x_i)$ with  $Y =
            (\pub(x_1)\times\dots\times\pub(x_k))$}
    \end{align*}
    }
    \parbox[b]{.4\textwidth}{
    \begin{align*}
        \first{\an{x,y}} &= x  
            &\second{\an{x,y}} &= y \\[2ex]
        x \doteq x  \quad &= \quad \True 
        &\True \vee x \quad &= \quad \True \\[1ex]
        \True \wedge \True  \quad &= \quad \True 
        & x \vee \True  \quad &= \quad \True 
    \end{align*}
    }
    \caption{Theory $E$ --- equational theory for modeling Civitas.\label{fig:eqth}}
    \end{figure*}
     
    The equational theory for modeling these primitives is given in
    the appendix (Fig.~\ref{fig:eqth}). We assume additionally that
    $+$ and $\times$ are equipped with equations for associativity and
    commutativity property (we could consider more complex equational
    theory for there operators, which  however makes the proof more
    complicated).  This theory will be denoted by $E$. 

    \subsection{Zero-knowledge Proofs}

    We will model the zero-knowledge proofs used in the protocol
    following the approach of \cite{BackesMaffeiUnruh-SP-2008}.  A
    zero-knowledge proof will be represented by a term $P =
    \zk^{n,k}_\phi(t_1,\dots,t_n; s_1,\dots,s_k)$ where
    $t_1,\dots,t_n$ are terms called \emph{the private component} (the
    proof will keep these terms secret), terms $s_1,\dots,s_k$ are
    called \emph{the public component} (the proof reveals these
    terms), and $\phi$ is a term built upon variables $x_1,\dots,x_n,
    y_1,\dots,y_n$ (no other variables and no nonces can occur in this
    term; $x_i$ is intended to refer to $t_i$, while $y_i$ is intended
    to refer to $s_i$), called \emph{the formula of $P$}.

    We have the following equalities associated to zero-knowledge
    proofs. The first group of equations reveals the public
    components (also the formula) of a proof. The second one allows
    one to check validity
    of a proof.
    \begin{align*}
        \public( \zk^{n,k}_\phi(t_1,\dots,t_n,s_1,\dots,s_k) ) 
            &= \an{\phi, s_1,\dots,s_k} \\[.4ex]
        \checkznp( \zk^{n,k}_\phi(t_1,\dots,t_n,s_1,\dots,s_k) ) 
            &= \True \quad\\
            \makebox[1cm]{
            \parbox[t]{20em}{\small if $\phi$ is a formula build upon
            $x_1,\dots,x_n,y_1,\dots,y_k$, and $\phi[t_i/x_i,s_i/y_i]
            \equiv_E \True$.}
            }
    \end{align*}

    To model Civitas, we will use zero-knowledge proofs formally
    defined in Fig.~\ref{fig:zk}. We use semicolons only to enhance
    legibility, as a mean of separating private and public components.
    The meaning of these proofs is as follows.
    \begin{figure*}
        \begin{align*}
            \knowPK  \text{ stands for }  &
                \zk^{1,1}_\phi &&\text{with }
                \phi = \bigl( y_1 \doteq \pub(x_1) \bigr) \\
            \dvpr \text{ stands for }  &
                \zk^{2,4}_\phi &&\text{with } \phi = \bigl( y_2 \doteq
                \reenc(y_1,y_3,x_2) \vee y_4\doteq\pub(x_2) \bigr) \\
            \proofdpart \text{ stands for }  &
                \zk^{1,3}_\phi &&\text{with } \phi = \bigl(
                y_1\doteq\dpart(y_3,x_1) \wedge y_2\doteq \pub(x_1)
                \bigr) \\
            \proofpetpart \text{ stands for }  &
                \zk^{1,4}_\phi &&\text{with } \phi = \bigl(
                y_1\doteq\petpart(y_3,y_4,x_1) \wedge y_2\doteq \pub(x_1)
                \bigr) \\
            \proofvote_l \text{ stands for } &
                \zk^{1,3}_\phi  &&\text{with } \textstyle
                \phi = \bigvee_{i=1}^l(y_1 \doteq \parenc{y_3[i]}{y_2}{x_1}) \\
            \mutknow  \text{ stands for }  &
                \zk^{2,5}_\phi &&\text{with }
                \phi = \bigl(y_1\doteq\parenc{x_1}{y_3}{x_3} \wedge
                y_2\doteq\parenc{x_2}{y_3}{x_4} \bigr) \\
            \proofmix_l \text{ stands for }  &
                \zk^{1,3}_\phi &&\text{with } \textstyle
                \phi = \bigvee_{\pi\in P_l}
                \bigwedge_{i=1}^l \bigl( y_2[\pi(i)] \doteq
                \reenc(y_1[i],y_3,x_1[i])  \bigr) 
        \end{align*}
        \caption{Shortcuts for zero-knowledge proofs.\label{fig:zk}
        In the equations $t[i]$ denotes the $i$-th element of a tuple
        $t$ (obtained by appropriately applying destructors to $t$),
        and $P_l$ denotes the set of all permutation
        of $\{1,\dots,l\}$.}
    \end{figure*}

    \begin{enumerate}
    \item[$\knowPK(x;\;y)$] 
        represents a proof of knowledge of the private key $x$
        associated with the given public key $y$ (i.e.\ $y=\pub(x)$).

    \item[$\dvpr(\alpha,x;\; m,m',k, k_v)$]
        represents a designated-verifier reencryption proof which
        shows  that $m'$ is a reencryption of $m$ under $k$; $k_v$ is
        the public key of the designated verifier who, having the
        corresponding private key, is able to forge a faked proof;
        $\alpha$ is an additional randomness used to construct the
        proof.  The proof is valid if either (a) $m' = \reenc(m,k,x)$
        or (b) $k_v=\pub(x)$, i.e.\ $x$ is a private key associated
        with public key $k_v$ of the designated verifier.
        
    \item[$\proofdpart(x;\; p,y,c)$]
        represents a proof that $p$ is the public share for
        distributed decryption of $c$ w.r.t.\ $y$, i.e.\  $p =
        \dpart(c,x)$ and $y=\pub(x)$.

    \item[$\proofpetpart(x;\; p,y,c,c')$]
        represents a proof that $p$ is the public share for
        distributed PET of ciphertexts $c$ and $c'$ w.r.t.\ $y$, i.e.\
        $p = \petpart(c,c',x)$ and
        $y=\pub(x)$.

    \item[$\proofvote_l(r;\;m,k,\vec b)$]
        represents a proof that $m$ is an encryption under $k$ of one
        of the values in $\vec b = \an{b_1,\dots,b_l}$ ($m =
        \parenc{b}kr$, where $b$ is an element of $\vec b$).

    \item[$\mutknow(m,m',r,r';\; c,c',k)$]
        represents a proof of mutual knowledge of the plaintexts
        contained in ciphertexts $c$ and $c'$ ($c = \parenc{m}{k}{r}$
        and $c' = \parenc{m'}{k}{r'}$).
        
    \item[$\proofmix_l(\vec r;\; \vec c_1, \vec c_2, k) $]
        where $\vec r, \vec c_1, \vec c_2$ are tuples of length
        $l$, represents a proof that $\vec c_2$ is obtained from
        ciphertexts $\vec c_1$ by mixing (i.e.\ applying some
        permutation) and reencryption ($\vec r$ is the collection of
        random values used in reencryption), i.e.\ $\vec c_2[\pi(i)] =
        \reenc(\vec c_1[i],k,\vec r[i])$, for some permutation $\pi$ of
        $\{1,\dots,l\}$.    
    \end{enumerate}

\subsection{Protocol Description}\label{sect:civitas:descr}

    \paragraph*{The participants.} The participants of the protocol
    are: the voters $\voter_0,\dots,\voter_m$, the supervisor $\auts$,
    the bulletin board $\autb$ registration tellers $\autr_0,\dots,
    \autr_k$, ballot boxes $\autbox_0, \dots, \autbox_k$, and
    tabulation tellers $\autt_0,\dots,\autt_k$.  We will assume that
    $\autb$, $\autr_0$, $\autbox_0$, and $\autt_0$ are honest. The
    remaining voting authorities may be dishonest. We will also assume
    that some of voters are dishonest and cooperate with the coercer.
    We assume that the channel from the voter's trusted registration
    teller is untappable. 

    \medskip\noindent
    In what follows, we assume that $i$ ranges over the set
    $\{0,\dots,m\}$ and $j$ ranges over $\{0,\dots,k\}$. For a
    participant $a$, we will write $\sig ma$ instead of $\sig
    m{\pub(k_a)}$. We will also write $\pub(a)$ instead of
    $\pub(k_a)$. 

    \paragraph*{Setup phase.}
    We do not model here the first part of the setup phase, where the
    supervisor posts the ballot design (the set of valid votes),
    identifies the tellers by posting their public keys, and posts the
    electoral roll (the set of authorized voters). Instead, we assume
    that the public keys of the voting authorities, the ballot design,
    and the electoral roll are fixed. Below, we describe the remaining
    steps of this phase.

    Tabulation tellers collectively generate a public key for a
    distributed encryption scheme and post it on the bulletin board
    (decryption of messages encrypted under this key requires the
    participation of all tabulation tellers):
    \begin{protocol}
        \mstep{KGen1}{\autt_j}{\autb}
        { 
            $\sig{h(y_j)}{\autt_j}$
        }
        \mstep{KGen2}{\autt_j}{\autb}
        { 
            
            $\sig{y_j, \; \knowPK(x_j; y_j)}{\autt_j}$
        }
    \end{protocol}
    where $y_i = \pub(x_i)$ and $x_i$ is random value, the private key
    share of $\autt_j$.  After the first step, all the tellers wait
    until all commitments are available. After the second step, they
    check proofs (by ``checking a proof $p$'' we mean verifying that
    its public components are as required and that
    $\checkznp(p)=\True$).  Now, $(y_1\times\dots\times y_k)$ is the
    distributed public key of $\autt_1,\dots,\autt_k$. We will refer
    to this key by $\keyt$.
        
    Next, each registration teller $\autr_j$ randomly generates
    \emph{credential} shares $s_{ij}$ (for each voter $\voter_i$)
    and post these shares on the bulletin board:
    \begin{protocol}
        \mstep{Cred}{\autr_j}{\autb}
        { 
            $\sig{i,\, S_{ij}}{\autr_j}$ \quad (for each $i,j$)
        }
    \end{protocol}
    where $S_{ij} = \parenc{s_{ij}}{\keyt}{r_{ij}}$, and $r_{ij}$ are
    random. The \emph{public credential of $\voter_i$} is now publicly
    computable as $S_i = (S_{i1}\times\dots\times S_{ik})$. 

    \paragraph*{Registration phase.}
    Voters register to acquire their private credentials:
    \begin{protocol}
        \mstep{Reg1}{\voter_i}{\autr_j}
        { 
            $\request$
        }
        \mstep{Reg2}{\autr_j}{\voter_i}
        { 
            $s_{ij},\; \bar r_{ij},\; D_{ij}$
        }
    \end{protocol}
    where $\bar r_{ij} = (r_{ij}+w_{ij})$, for random $w_{ij}$ and 
    $D_{ij} = \dvpr(\delta_{ij}, w_{ij};\; S_{ij}, S'_{ij}, \keyt,
    \pub(\voter_i))$, with $S'_{ij} = \parenc{s_{ij}}{\keyt}{\bar
    r_{ij}}$ (which, up to the equation theory under consideration,
    equal to $\reenc(S_{ij}, \keyt, w_{ij})$) and random
    $\delta_{ij}$, is a designated-verifier reencryption proof which
    shows that $S'_{ij}$ is a reencryption of $S_{ij}$. The voter
    verifies this proof.  Now, his private credential is $s_i =
    (s_{i0}\times \dots\times s_{ik})$.

    \paragraph*{Voting phase.} Each voter sends his \emph{ballot} $b_i$
    containing his vote along with his credential to all ballot boxes:
    \begin{protocol}
        \mstep{Vote}{\voter_i}{\autbox_j}
        { 
            $b_i = \an{ \parenc{s_i}{\keyt}{r_i}, \,
             \parenc{v_i}{\keyt}{r'_i},\, P_V^i,\, P_K^i }$ 
        }
    \end{protocol}
    where $r_i,r'_i$ are random, $v_i$ is the vote chosen by
    $\voter_i$, $P_V^i = \proofvote_l(r'_i; \parenc{v_i}{\keyt}{r'_i},
    \keyt, b_1,\dots,b_l )$, and $P_K^i = \mutknow(s_i, v_i, r_i,
    r_i'; \parenc{s_i}{\keyt}{r_i}, \parenc{v_i}{\keyt}{r'_i},
    \keyt)$.  The value $v_i$  will be called \emph{the vote of
    $b_i$}; $s_i$ is \emph{ballot credential} of $b_i$;
    $\parenc{v_i}{\keyt}{r'_i}$ will be called \emph{the encrypted
    vote of $b_i$}, and $\parenc{s_i}{\keyt}{r_i}$ will be called
    \emph{the encrypted credential of $b_i$}.  $P_V^i$ is a
    zero-knowledge proof which shows that the vote is well-formed with
    respect to the ballot design ($v_i$ is one of the valid votes
    $z_1,\dots,z_l$), and $P_K^i$  is a zero knowledge-proof which
    shows that the submitter simultaneously knows $s_i$ and $v_i$.  We
    will some times write $b_i[v',s']$ for the message like $b_i$ but
    with $v'$ and $s'$ instead of $v_i$ and $s_i$.  

    \paragraph*{Tabulation phase.}
    Before the tabulation phase, each ballot box posts a commitment to
    its contents on the bulletin board:
    \begin{protocol}
        \mstep{Comm1}{\autbox_j}{\autb} 
            { $\sig{j,C_j}{\autbox_j}$ }
    \end{protocol}
    where $C_j = h(\mathrm{contents}(\autbox_j))$.
    The supervisor then posts his own signatures on all these
    commitments, defining the set of votes to be tabulated:
    \begin{protocol}
        \mstep{Comm2}{\auts}{\autb} 
            { $\sig{j,C_j}{\auts}$ }
    \end{protocol}

    Then, the tabulation tellers collectively tally the election: All
    tabulation tellers (1) \emph{retrieve} the ballots from all
    ballot boxes and the public credentials from the bulletin board.
    They also verify that the content of ballot boxes corresponds to
    the commitments posted in (Comm2). Then, they (2) \emph{check proofs}
    in retrieved ballots and eliminate any ballot with an invalid proof.
    Note that these steps are performed by each teller independently,
    and the resulting set of votes, let us denote it by $B$, is
    determined by the publicly known information.

    Next, (3) \emph{duplicate elimination} (according to some fixed
    policy) is performed, by running $\pet(c,c')$, for all
    encrypted ballot credentials $c,c'$ from distinct ballots in $B$:
    \begin{protocol}
        \mstep{PET1}{\autt_j}{\autb} 
            { $\sig{\alpha_j(c,c'),\; P_{\alpha_j}(c,c')}{\autt_j}$ }
    \end{protocol}
    where 
    $\alpha_j(c,c') = \petpart(c,c',x_j)$ and
    $P_{\alpha_j}(c,c') = \proofpetpart(x_j;  \alpha_j(c,c'), y_j, c,
    c')$.
    Now, each teller waits until all the tellers post their share
    and verifies the proofs. The result of PET for $c,c'$ is
    $\distpet(\alpha_0,\dots,\alpha_k)$ (it evaluates to $\True$ if the PET
    passes) and is publicly computable. For each two ballots for which
    PET holds true, only one is kept (according to the mentioned
    policy).

    Next, (4) \emph{mixing ballots} is performed on the list of
    remaining ballots $\vec u_0$.  Each tabulation teller in turn
    applies its own random permutation $\pi^b_j$ with reencryption.
    We assume that $\vec u_j$ is the input for $j$-th teller:
    \begin{protocol}
        \mstep{Mix1}{\autt_j}{\autb} { 
            $\sig{\vec u_{j+1},\; P_{u_j}}{\autt_j}$
        }
    \end{protocol}
    where $\vec r_j$ is a vector of random values, $\vec
    u_{j+1}[\pi^b_j(i)] = \reenc(\vec u_j[i], \keyt, \vec r[i])$, and
    $P_{u_j} = \proofmix(\vec r_j;\; \vec u_j,\, \vec u_{j+1},\,
    \keyt)$. The result of mixing is $\vec u_{k+1}$.  Similarly,
    \emph{mixing credentials} is performed on the list $\vec w_0 =
    (S_0,\dots,S_m)$ of public credentials.  Each tabulation
    teller in turn applies its own random permutation $\pi^c_j$ with
    reencryption.  We assume that $\vec w_j$ is the input for $j$-th
    teller: 
    \begin{protocol}
        \mstep{Mix2}{\autt_j}{\autb} { 
            $\sig{\vec w_{j+1}, \; P_{w_j}}{\autt_j}$
        }
    \end{protocol}
    where $\vec r'_j$ is a sequence of random values, $\vec
    w_{j+1}[\pi^c_j(i)] = \reenc(\vec w_j[i], \keyt, \vec r'[i])$, and
    $P_{w_j} = \proofmix(\vec r'_j;\; \vec w_j,\, \vec w_{j+1},\,
    \keyt)$. The result of mixing is $\vec w_{k+1}$.

    The next step is \emph{invalid ballots elimination} where ballots without
    valid credentials are eliminated. For each ballot with the
    encrypted credential $c$, $\pet(c,c')$ is performed against every
    public credential $c'$:
    \begin{protocol}
        \mstep{PET2}{\autt_j}{\autb} 
            { $\sig{\beta_j(c,c'),\; P_{\beta_j}(c,c')}{\autt_j}$ }
    \end{protocol}
    where $\beta_j(c,c') = \petpart(c,c',x_j)$ and $P_{\beta_j}(c,c')
    = \proofpetpart(x_j;\; \beta_j(c,c'), y_j, c, c')$.  Now, each
    teller waits until all the tellers post their share and verifies
    the proofs. The result of PET for $c,c'$ is
    $\distpet(\beta_0,\dots,\beta_k)$ and is publicly computable.  If
    this test fails for all $c'$, the ballot is removed.

    Finally, \emph{decrypt} step is performed, for each of the
    remaining ballots. Decryption is applied to the encrypted vote
    $c$ of each of the remaining ballots (but not to the encrypted
    credentials): 
    \begin{protocol}
        \mstep{Decr}{\autt_j}{\autb} { 
            $\sig{\gamma_j(c),\; P_{\gamma_j}(c)}{\autt_j}$
        }
    \end{protocol}
    where $\gamma_j(c) = \dpart(c,x_j)$ and $P_{\gamma_j}(c) =
    \proofdpart(x_j;\; \gamma_j(c), y_j, c)$.
    Each teller waits until the remaining tellers submit their shares
    and verifies the proofs. Now, the decrypted vote is $v =
    \distdec(\gamma_0,\dots,\gamma_k)$. At this point the result of the voting
    process is publicly computable.

\subsection{Modelling of the Protocol}    
    
    In addition to the participants enumerated in Section
    \ref{sect:civitas:descr}, we assume that the coercer $\coercer$
    and a key issuer $\autk$ also participate in the protocol. The
    role of the key issuer is to generate private and public keys to each
    participant and, to provide these keys on request.
    
    We assume that $\voter_0$ is the coerced voter.
    The voters $\voter_1,\dots,\voter_n$, for some $n<m$, are honest,
    while $\voter_{n+1},\dots,\voter_m$ are dishonest and will not be
    modelled directly, but, instead, will be subsumed by the coercer.
    As we mentioned before, we assume that $\auts$, $\autb$,
    $\autr_0$, $\autbox_0$, and $\autt_0$ are honest.
    The remaining authorities, that is  $\autbox_i$, $\autr_i$,
    $\autt_i$, for $i\in\{1,\dots,k\}$, are assumed to be dishonest
    and will be subsumed by the coercer. Additionally, to model
    anonymous channels, we introduce agents $\anch_0,\dots,\anch_k$.
    The role of $\anch_k$ is to simply forward messages to $\autbox_i$
    (and so, $\autbox_j$ cannot associate a sender to a received
    message).
   
    The set $\channels$ of the channels used in the protocol consists
    of: 
    \begin{compactitem}[\hspace\parindent--] 
    \item
        $\ch\auts x$, for each protocol participants $x$ (including
        $\auts$),
    \item
        $\ch{\autk}{x}$ and $\ch{x}{\autk}$, for each protocol
        participant $x\neq\autk$,
    \item
        $\ch{\autb}{x}$ and $\ch{x}{\autb}$, for each protocol
        participant $x\neq\autb$, $x\neq\anch_j$, and $x\neq\autbox_j$,
    \item
        $\ch{\anch_j}{\autbox_j}$ and
        $\ch{x}{\anch_j}$, for each protocol participant $x$,
    \item
        $\ch{\autbox_j}{\autb}$ 
    \item
        $\ch{\voter_i}{\autr_j}$ and $\ch{\autr_j}{\voter_i}$,
    \item
        $\ch{\voter_0}{\coercer}$ and $\ch{\coercer}{\voter_0}$,
    \item 
        $\ch{init}{\auts}$ is a channels used to initiate $\auts$.
    \end{compactitem}
    We will use the following notation: for a set $B$, we will write
    $\ch B*$ for the set of all $c\in\channels$ of the form $\ch xy$
    with $x\in B$. Similarly, we will write $\ch *B$ for the set of all
    $c\in\channels$ of the form $\ch xy$ with $y\in B$. 

    We assume that $\Sigma$ contains, in addition, constants $\init$,
    $\request$, and $\done$ and also constants representing the participant
    names.

    The protocol we take is $S = (A,\vin,\vout,s_0,P)$, where $A =
    \{\voter, \coercer, \env\}$ and $\vin$, $\vout$, $s_0$, and $P$
    are defined as follows: $\vin$ and $\vout$ are the functions:
    \begin{align*}
        \vin(\voter) &= \ch{*}{v_0} 
          &\vout(\voter) &= \ch{v_0}* \\
        \vin(\coercer) &= \ch{*}{D}
          &\vout(\coercer) &= \ch{D}* \\
        \vin(\env) &= \ch{*}{H}
          &\vout(\env) &= \ch{H}*
    \end{align*}
    where $H = \{\auts, \autb, \autr_0, \autbox_0, \autt_0,
    \voter_{1}, \dots, \voter_n\}$ is the set of honest participants
    and $D = \{\coercer, v_{n+1}, \dots, v_m, \autr_1, \dots, \autr_k,
    \autbox_1, \dots, \autbox_k, \autt_1,\dots, \autt_k \}$ is the set
    of dishonest participants.  Additionally, both
    $\voter_0$ and $\coercer$ have an infinite number of private
    channels (i.e.\ channels that occur only in $\vin(\voter) \cap
    \vout(\voter)$ or $\vin(\coercer) \cap \vout(\coercer)$,
    respectively). In particular, let $c_v$ be some private channel of
    $\voter$.

    The \emph{initial sequence} $s_0 = (\overline{\ch{init}{\auts}}:
    \init)$. For each participant $a$, we define the set $P(a)$ of
    programs of this participant as follows.
    
    \paragraph*{\it Key-Issuer.}
    $P(\autk)$ consists of programs which assign a distinct nonce
    $k_a$ (the private key of $a$) to each participant $a$ and, in
    response to $(\ch{x}{\autk}:\request)$,
    send on channel $\ch\autk x$ the tuple containing the public keys
    $\pub(k_a)$ of all the participants and, additionally, the private
    key $k_x$ of $x$. 

    \paragraph*{\it Bulletin board.}
    The set $P(\autb)$  contains one program which immediately
    forwards all received messages to all participants (except for
    $\autbox_j$).

    \paragraph*{\it Ballot box.}
    The set $P(\autbox_0)$  contains one program which, after obtaining
    the message $\done$ from $\auts$, posts the commitment of its
    content (i.e.\ the list of the messages received so far) on
    the bulletin board and publishes this content, i.e.\ sends the
    content to all participants (except for $\autb$, $\anch_j$, and
    $\autbox_j$).  In addition, this program immediately forwards to
    the coercer each message he receives (this model the fact that the
    coercer is able to intercept messages sent to the ballot box, but
    cannot block them).

    \paragraph*{\it Supervisor.}
    The set $P(\auts)$ consists of a program which, in
    response to the message $\init$, initiates $\autt_j$ and $\autr_j$
    by sending them $\init$ message and waits until the setup phase is
    completed (all the necessary commitments and key shares are
    posted). Then it sends  $\init$ to all voters and message $\done$
    to itself (this models ``waiting'' for the voting phase to end).
    When this message is delivered, it sends $\done$ to all the
    ballot boxes and waits for their commitments. After it obtains
    these commitments, he signs them and posts on the bulletin board.

    \paragraph*{\it Registrars.}
    The set $P(\autr_0)$ consists of programs which,
    in response to the message $\init$ sent by $\auts$, request for
    keys and, after obtaining them, pick a distinct nonces $s_{ij}$
    and $r_{ij}$, and post $\sig{\voter_i,S_{ij}}{\autr_j}$ to the
    bulletin board, as defined in (Cred), for each
    $i\in\{0,\dots,m\}$. Then, on request sent by $\voter_i$, it
    replies with (Reg2).
     
    \paragraph*{\it Honest voters.}
    The set $P(\voter_i)$, for $i\in\{1,\dots,n\}$, consists of (a)
    programs $\xi_z$, for each valid vote $z$, which after receiving
    message $\init$ from $\auts$, take the keys from $\autk$, request
    for credentials (Reg1) and, after obtaining them all (Reg2), post
    their ballots $b_i$ (Vote), with $v_i= z$ and fresh nonces $r_i$
    and $r_i'$, to all $\anch_j$; (b) programs which register like
    $\xi_z$ but do not post any ballot (abstain from voting); (c)
    programs $\xi_\bot$, which is defined like $\xi_z$, but instead of
    posting a valid ballot, posts a ballot with an invalid credential
    (some fresh nonce);
    
    \paragraph*{\it Anonymous channel.}
    The set $P(\anch_j)$ consists of one program
    which forwards to $\autbox_j$ every message it receives.

    \paragraph*{\it Tallier.} The set $P(\autt_0)$ consists of the
    following programs: a program, after receiving $\init$ from
    $\auts$, participates in the procedure of public key generation: it
    picks a nonce $x_i$ (its private key) and posts (KGen1) and
    then, when it sees that all the tellers have posted their messages
    (note that he can see it, because the bulletin board forwards all
    the messages to every participant), it post (KGen2), waits for the
    corresponding messages of the remaining tellers and checks the
    proofs (it these tests fail, it halts).

    Then, after it obtains (forwarded by the bulletin board) the commitment
    of $\auts$ on the contents of all ballot boxes (sent in step
    (Comm2)), it participates in the tabulation procedure:
    it post messages as defined in steps (PET1)--(Decr). After each
    step, it waits for the remaining tellers to post their messages
    and verifies whether these messages have an appropriate form and
    the zero-knowledge proofs are correct. If these tests fail, it halts.

    \paragraph*{\it Coerced voter.}
    The set $P(\voter_0)$ consists of program of the form $(\vreg
    \comp v)$, where
    $v\in\Proc(I_0,O_0)$, for $O_0=\vout(\voter_0)$ and $I_0 =
    \vin(\voter_0) \setminus \{\ch{\auts}{\voter_0},
    \ch{\autk}{\voter_0}, \ch{\autr_j}{\voter_0}\}$, and $\vreg$ is the
    program which after receiving message $\init$ from $\auts$, take
    the keys from $\autk$, request for credentials (Reg1) and, after
    obtaining them all, sends all the obtained keys and credentials on
    $c_v$. The program $\vreg$, performing registration, is the fixed
    part of any program of $\voter_0$; $v$ represents the behaviour of
    $\voter_0$ after registration has been done (e.g.\ in the voting
    phase). Note that $v$ has access to all
    the registration data, as can read the data sent on $c_v$. 

    \paragraph*{\it Coercer.}
    $P(\coercer)$ is defined as $\Proc(\vin(\coercer),\vout(\coercer))$.  

    \subsection{Proof of Theorem~\ref{th:civitas}}

    Let us denote Civitas with restricted coercion strategies by $S$.
    Note that this protocol is not normal, because the set $V$ does
    not contain all programs over $\vin(\voter)$, $\vout(\voter)$, and
    thus we cannot use Theorem~\ref{th:dummy}. Hence, we first show
    that coercion-resistance of this protocols is equivalent to
    coercion resistance of some normal protocol $\hat S$.  Let $\hat
    S$ be defined like protocol $S$ with only one difference: the
    subprocess $\vreg$ of any program of the coerced voter will be now
    run by a distinct agent $\voter^*_0$, which will be a part of the
    environment. 
    
    Let $R$ be the coercion system induced by $S$, and $\hat R$ be the
    coercion system induced by $\hat S$. Let $\rho$ be a run of $S$,
    which means that $\rho = ((\vreg\comp v),c,e,\pi)$, where $\pi$ is
    induced by $(\vreg \comp v \comp c \comp e)$. By $\hat\rho$ we
    denote $(v, c, (\vreg \comp e), \pi)$. Note that $\hat\rho$ is a
    run of $\hat S$. We extend the operator $\hat\cdot$ to properties
    of $S$ in a natural way: 
    $\hat\beta = \{\hat\rho : \rho\in\beta\}$.  It is easy to show
    that the following lemma holds.

    \begin{lemma}
        $R$ is coercion-resistant in $\alpha$ w.r.t.\ $\gamma_z$ iff
        $\hat R$ is coercion-resistant in $\hat\alpha$ w.r.t.\
        $\hat\gamma_z$.
    \end{lemma}

    One can show that protocol $\hat S$ is normal and both
    $\alpha$ and $\gamma$ are now closed under $\peq$. Hence, we can use
    Theorem~\ref{th:dummy}. So, it is enough to provide a
    counter-strategy $v'$ for a strategy $v$ which simply forwards to
    the coercer all the messages obtained from the remaining
    participants and forwards to these participants all the messages
    obtained from the coercer.
    
    Let $v'$ be the process which after obtaining the registration
    data on $c_v$ post the ballot $b_0 = b[z,s_0]$ and, in the same
    time, behaves like the forwarder $v$ with the following exception.
    When he obtains the registration data on $c_v$, he changes it
    before forwarding: he replaces $s_{00}$ by a fresh nonce $\tilde
    s_{00}$ (a faked credential) and $D_{00}$ by $\tilde D_{00} =
    \dvpr(\tilde\delta_{00}, k_{\voter_0};\, S_{00}, \tilde S'_{00},
    \keyt, \pub(\voter_0))$ (a faked proof) with random
    $\tilde\delta_{00}$ and $\tilde S'_{00} = \parenc{\tilde
    s_{00}}{\keyt}{\bar r_{00}}$ (recall that $\pub(\voter_0)$ stands
    for $\pub(k_{\voter_0})$).  We will show that $v'$ is a
    counter-strategy for $v$.

    \medskip \textbf{First, we show that condition (iii) of
    Definition~\ref{def:cr} holds} for $v'$. Let $\rho$ be a run of
    the system induced by $v'$, i.e.  $\rho$ is induced by $(v' \comp
    c \comp e)$, for some $c\in C, e\in E$.  If $\rho$ is not fair,
    then there is nothing to prove.  So, suppose that $\rho$ is fair.
    First, note that, by the fairness assumption, all $R_0,\dots,R_k$
    post all messages and zero-knowledge proofs as required. Since
    $\vreg$ does not send out his private key, the DVRP-s he gets
    cannot be faked, and thus the private credential he obtains is
    valid. Second, $v'$ obtains his registration data before the
    voting phase ends\footnote{By the expression ``$\voter_0$ obtains his
    credentials'', used in the definition of a fair run w.r.t.\
    $\voter_0$, we mean formally that this credential is delivered to
    $v'$.}.
    
    Since $v'$ posts then a valid ballot $b_0$ right
    away (still before the voting phase ends), by the fairness
    assumption, this ballot is posted successfully and so $b_0$ is in
    the initial pool of votes to be tabulated. 
    
    Now, it is easy to show that $b_0$ will be successfully processed
    by tabulation tellers, using the fact that $v'$ never reveals his
    private credential (so it is not used in any other ballot) and
    the assumption that the run is fair (and so all the tabulation
    tellers have to correctly perform all the expected step, because
    otherwise they would not be able to construct valid zero-knowledge
    proofs).

    \medskip \textbf{Now, we will show that condition (i) of
    Definition~\ref{def:cr} holds} for $v$ and $v'$. So, let
    $\rho\in\alpha$ be a run induced by $(v \comp c \comp e)$, for
    some $c\in C$ and $e \in E$.

    Since $\rho$ is in $\alpha$, there is some honest voter, say
    $\voter_1$, who successfully posts a ballot $b_1[z,s_1]$
    (that is a ballot with vote $z$), some honest voter, say
    $\voter_2$, who obtains his credential and successfully
    posts a ballot $b_2[z_2,s^*_2]$ with an invalid credential
    (a fresh nonce $s^*_2$), and some honest voter, say
    $\voter_3$, who posts his ballot $b_3$ after $\voter_0$
    finishes registration.
    
    We take $e'\in E$ which is like $e$ with the following
    exceptions: (a) $\voter_3$ abstains from voting, (b) if
    $\voter_3$ in $\rho$ posted his ballot successfully, then
    $\voter_1$ votes like $\voter_3$ voted in $\rho$; and (c)
    moreover, if at least one proper ballot with $s_0$ is in
    $\rho$ successfully posted and $z_c$ is the vote in the
    ballot with $s_0$ that is kept after duplicate elimination
    (note that $z_c$ must be a valid vote), then $\voter_2$
    posts a valid ballot with $z_c$ instead of the invalid one.
    Also, instead of using permutations $\pi^b_0$ and $\pi^c_0$
    in steps (Mix1) and (Mix2), $\autt_0$ uses slightly
    different permutations (see Sect.
    \ref{sect:civitas-details}).

    The run $\rho'$ of $(v' \comp c \comp e')$ is constructed from
    $\rho$ in the following way.  The messages in $\rho'$ are
    delivered in the same order like the corresponding messages in
    $\rho$ with the following exceptions: first, the message sent by
    $\vreg$ on $c_v$ is delivered immediately and, second,
    the ballot sent by $\voter_0$ in $\rho'$ is delivered at the same
    step, when the ballot sent by $\voter_3$ is delivered in $\rho$
    (because $\voter_0$ sends his ballot, just when he gets the
    registration data on $c_v$, and $\voter_3$ posts his ballot after
    it, the ballot of $\voter_0$ is ready to be delivered at the
    mentioned step).
    
    Now, one can show that $\rho \sim \rho'$. The rough idea is as
    follows: $\voter_2$ is used to hide the fact that valid ballots
    possibly posted by the coercer in $\rho$ become invalid in
    $\rho'$, $\voter_3$ is used to hide the fact that $\voter_0$ posts
    his ballot in $\rho'$, but abstains from voting in $\rho$, and
    finally $\voter_1$ is used to balance the outcome of the voting.
    Due to the fact that the coercer cannot tell any difference
    between an original DVRP and a faked one and the fact that the
    messages posted on the bulletin board are mixed and reencrypted
    before decryption, the frames are indistinguishable to the
    coercer.

    Details of the proof depend on (a) whether or not the coercer
    successfully posts at least one valid ballot with $s_0$ and (b)
    whether or not $\voter_3$ posts his ballot successfully.  In next
    subsection, we present a detailed proof for one of these cases:
    when the coercer successfully posts one proper ballot with $s_0$,
    and $\voter_3$ also posts his ballot successfully. 

    \medskip \textbf{Now, we will show that condition (ii) of
    Definition~\ref{def:cr} holds} for $v$ and $v'$. So, let
    $\rho\in\alpha$ be a run induced by $(v' \comp c \comp e)$, for
    some $c\in C$ and $e \in E$. We define the vote $z_c$ as
    follows: if the coercer, in $\rho$, successfully posts at least
    one ballot with $\tilde s_0$ (where $\tilde s_0$ is computed
    like $s_0$, but using $\tilde s_{00}$ instead of $s_{00}$),
    then let $z_c$ be the vote in the ballot containing $\tilde
    s_0$ which is left after duplicate elimination phase;
    otherwise let $z_c$ by any vote. Note that, since a valid
    ballot has to contain a proof that a vote in it is valid,
    $z_c$ must be a valid vote.

    Since $\rho$ is in $\alpha$, there is some honest voter, say
    $\voter_1$, who obtains his credential, before $\vreg$ finishes
    registration, but abstains from voting, and some honest voter, say
    $\voter_2$, who successfully votes for $z_c$.

    We take $e'\in E$ which is as $e$ with the following
    exceptions: $\voter_1$ votes for $z$ and, moreover, if at
    least one ballot with $\tilde s_0$ is, in $\rho$,
    successfully posted, then $\voter_2$ posts $b[z_c,s^*_2]$,
    for some unused nonce $s^*_2$, instead of $b[z_c,s_2]$.  We
    also need to slightly change the permutations used by
    $\autt_0$ in (Mix1) and (Mix2).

    The run $\rho'$ of $(v \comp c \comp e')$ is constructed from
    $\rho$ in the following way.  The messages in $\rho'$ are
    delivered in the same order like the corresponding messages in
    $\rho$ with the following exceptions: The ballot
    sent by $\voter_1$ is delivered in the same step, when the ballot
    sent by $\voter_0$ was delivered in $\rho$ (it is possible because
    this ballot is posted before $\vreg$ finishes registration).
    
    Now, one can show that $\rho \sim \rho'$. The rough idea is as
    follows: $\voter_2$ is used to hide the fact that the ballots
    involving the credential given by the coerced voter and possibly
    posted by the coercer are invalid in $\rho$ but valid in $\rho'$;
    $\voter_1$ is used to hide the fact that $\voter_0$ posts his
    ballot in $\rho$, but not in $\rho'$.  Moreover, due to the fact
    that the coercer cannot distinguish an original DVRP and the faked
    one, and the fact that the messages posted on the bulletin board
    are mixed and reencrypted, the frames are indistinguishable to the
    coercer.

    \subsection{Detailed Case Analysis}\label{sect:civitas-details}

    In this subsection we give a detailed proof that the runs $\rho$
    and $\rho'$, as constructed in the proof for condition (i) above,
    are indistinguishable to the coercer in the case the coercer, in
    $\rho$, successfully posts exactly one proper ballot
    $b_c[z_c,s_0]$, and $\voter_3$ also successfully posts his ballot
    $b_3[z_3,s_3]$. 
    
    Formally, we have to show the following.  The run $\rho$ is of the
    form $(\hat v,\hat c,\hat e,\pi)$, where $\hat v \simeq v$, $\hat
    c \simeq c$, $\hat e \simeq e$, and $\pi$ is a run induced by
    $(\hat v \comp \hat c \comp \hat e)$.  Similarly, $\rho'$ is of
    the form $(\hat v',\hat c,\hat e',\pi')$, where $\hat v' \simeq v'$,
    $\hat e' \simeq e'$, and $\pi'$ is a run induced by $(\hat v'
    \comp \hat c \comp \hat e')$. By the definition of $\sim$, we need
    to prove that $\pi \equiv_{\hat c} \pi'$, which mens that $\pi
    \equiv^{\nonc{\hat c}}_{\inp{\hat c}} \pi'$.  Since, $\inp{\hat c}
    \subseteq \vin(\coercer)$  it is enough to show that $\pi
    \equiv^N_I$ where $N = \nonc{\hat c}$ and $I = \vin(\coercer)$.
    This, by the definition of $\equiv^N_I$, is equivalent to the
    following statement, where $\phi = \pi_{|I}$ and $\phi' =
    \pi'_{|I}$: (i) $\chseq(\phi) = \chseq(\phi')$ and (ii) for each
    $\tau_1, \tau_2 \in T_N$, we have that $\tau_1[\phi] \equiv
    \tau_2[\phi]$ iff $\tau_1[\phi'] \equiv \tau_2[\phi']$.

    The proof goes as follows. First we describe $\phi$ and $\phi'$
    (which contain exactly those messages that are seen by the
    coercer) and show that condition (i) holds. Then we will show
    that condition (ii) holds as well.

    Let us first informally point out the differences in view of the
    coercer on $\rho$ and $\rho'$.  These views  are very similar, in
    particular, the lists of votes published by the tallying tellers
    in both cases are exactly the same.  The main differences are
    summarized in the table below, where $\tilde s_0$ denotes $\tilde
    s_{00} \times s_{01} \times\dots\times s_{0k}$ (i.e.\ the faked
    private credential of $\voter_0$).

    \begin{center} \small
        \begin{tabular}{lll} 
            & $\rho$ & $\rho'$ \\
        \hline \hline 
            the (faked) credential sent to $\coercer$ 
            & $s_0$ & $\tilde s_0$ \\ 
        \hline 
            the ballot posted by $\coercer$ 
             & $b_c[z_c, s_0]$ &  $b_c[z_c, \tilde s_0]$ \\ 
        \hline 
            the ballot posted by $\voter_3$\,/ $\voter_0$ 
            & $b_3[z_3, s_3]$ & $b_0[z, s_0]$ \\ 
        \hline
            the ballot posted by $\voter_1$ 
            & $b_1[z, s_1]$ & $b_1[z_3, s_1]$ \\ 
        \hline 
            the ballot posted by $\voter_2$ 
            & $b_2[z_2, s_2^*]$ & $b_2[z_c, s_2]$ \\ 
        \end{tabular} 
    \end{center}

    \smallskip\noindent
    The messages placed in the same raw of the table are seen by the
    coercer, in $\rho$ and $\rho'$, respectively, at the same channel
    and the same step. 

    We need also to specify the mentioned permutations $\tilde\pi^b_0$
    and $\tilde\pi^c_0$, used by $\autt_0$ in $\rho$, instead of
    $\pi^b_0$ and $\pi^c_0$, to mix votes and credentials. So,
    $\tilde\pi^b_0$ is like $\pi^b_0$, but places the reencryptions of
    $b_2[z_c,s_2]$, $b_1[z_3,s_1]$, $b_0[z,s_0]$, $b_c[z_c,\tilde
    s_0]$ in the place where $\pi^b_0$ places the reencryptions of
    $b_c[z_c, s_0]$, $b_3[z_3,s_3]$, $b_1[z, s_1]$, and $b_2[z_2,
    s^*_2]$, respectively.  The permutation $\tilde\pi^c_0$ is like
    $\pi^c_0$, but places the permutations of $\hat S_2$, $\hat S_1$,
    $\hat S_0$, $\hat S_3$ in the place where $\pi^c_0$ places the
    reencryptions of $\hat S_0$,
    $\hat S_3$, $\hat S_1$, $\hat S_2$, respectively, where $\hat S_i$
    are reencryptions public credentials produced in step (Mix2).

    Detailed description of $\phi$ and $\phi'$, which is
    the sequence of messages received by the coercer in $\rho$ and
    $\rho'$, respectively, is given in Fig.~\ref{fig:civitas-frames}.
        %
    \begin{figure*}[!t]\small 
    \[ \def\tl(#1){\textrm{\footnotesize(#1)}}
    \def\nl#1.{\textrm{\small(#1)}} 
    \begin{array}{r@{\ \ }l@{\ }r@{\qquad}l@{\ }l} 
    && \textrm{Messages in frame }\phi &
    \textrm{Messages in frame }\phi' \\[.1ex] 
    \hline \\[-.9ex]
        \nl1.&\tl(KGen1) & h(y_0) & h(y_0) \\[1ex] 
        \nl2.&\tl(KGen2) & y_0, \ \knowPK(x_0;\; y_0) & y_0, \
            \knowPK(x_0;\; y_0) \\[1ex]
        \nl3.&\tl(Cred) & i, \ S_{i0} & i, \ S_{i0} & \textrm{\footnotesize(for
            $i\in\{0..m\}$)} \\[1ex] 
        \nl4.&\tl(Reg2) & s_{i0}, \bar r_{i0}, D_{i0} & s_{i0}, \bar
            r_{i0}, D_{i0} & \textrm{\footnotesize(for $i\in\{n+1,\dots,m\}$)}\\[1ex] 
        \nl5.&\tl(Reg2) & s_{00}, \bar r_{00}, D_{00} & \tilde s_{00},
            \bar r_{00},  \tilde D_{00} \\[4ex]
        \nl6.&\tl(Vote) & {\mathbf B} = \left\{ \begin{array}{l}
        b_3[z_3,s_3], \ b_1[z, s_1],  \ b_2[z_2, s^*_2] \\ b_i
        \end{array}\right.\!\!\!\!\!  & \!\!\!\!\!\left.
        \begin{array}{r} b_0[z,s_0], \ b_1[z_3,s_1], \ b_2[z_c, s_2] \\ b_i
        \end{array}\right\} = \tilde{\mathbf B} & \textrm{\footnotesize(for 
        $i\in\{4..n$\})} \\[4ex]
        \nl7.&\tl(Comm) & \sig{j,C_j}{\autbox_j}, \ \sig{j,C_j}{\auts} &
        \sig{j,\tilde C_j}{\autbox_j}, \ \sig{j,\tilde C_j}{\auts}
        &\textrm{\footnotesize(for $j\in\{0,\dots,k\}$)} \\[2ex]
        \nl8.&\tl(PET1) & \alpha_0(c,c'),\ P_{\alpha_0}(c,c')\ \text{
        (for $c\neq c'$ in ${\mathbf B_c}$)} & \alpha_0(c,c'),\
        P_{\alpha_0}(c,c')\ \text{ (for $c\neq c'$ in $\tilde{\mathbf
        B}_c$)} \\[4ex]
        \nl9.&\tl(Mix1) & {\mathbf C} = \left\{ 
        \begin{array}{l} 
            \hat b_c[z_c, s_0], \, \hat b_3[z_3,s_3], \\ 
            \hat b_1[z, s_1] , \, \hat b_2[z_2, s^*_2] \\ 
            \hat b_i 
        \end{array}
        \right.\!\!\!\!\!  &
        \!\!\!\!\!\left. 
        \begin{array}{r} 
            \hat b_2[z_c,s_2], \ \hat b_1[z_3,s_1], \\ \hat b_0[z,s_0], \
            \hat b_c[z_c,\tilde s_0],  \\ \hat b_i 
        \end{array}
        \right\} = \tilde{\mathbf C} &\textrm{\footnotesize(for 
        $i=4..m$)}  \\[2ex] && P_{u_0} & \tilde
        P_{u_0} \\[4ex]
        \nl10.&\tl(Mix2) & {\mathbf S} = \left\{ \begin{array}{l} \hat
        S_0, \hat S_3, \hat S_1, \hat S_2 \\ \hat S_i
        \end{array}\right.\!\!\!\!\!  & \!\!\!\!\!\left.
        \begin{array}{r} \hat S_2, \hat S_1, \hat S_0, \hat S_3 \\
        \hat S_i \end{array}\right\} = \tilde{\mathbf S}
        &\textrm{\footnotesize(for
        $i=4...m$)} \\[2ex] && P_{w_0} & \tilde
        P_{w_0} \\[4ex]
        \nl11.&\tl(PET2) & \beta_0(c,c'), \ P_{\beta_0}(c,c') \quad\qquad\ 
        &
        \beta_0(c,c'), \ P_{\beta_0}(c,c') \\
        &&
          \textrm{\footnotesize(for $c\in {\mathbf C_c}$ and $c'\in {\mathbf S}$)}
        & \quad\
        \textrm{\footnotesize(for $c\in \tilde{\mathbf C}$ and $c'\in \tilde{\mathbf S}$)} \\[2ex]
        \nl12.&\tl(Decr) & \gamma_0(c), \ P_{\gamma_0}(c) \
        \textrm{\footnotesize
        (for $c\in {\mathbf C_v}$)} & \gamma_0(c), \ P_{\gamma_0}(c) \
        \textrm{ (for $c\in \tilde{\mathbf C_v}$)} \\[.8ex] \hline
        \end{array} 
        \] 
    \caption{\label{fig:civitas-frames} 
        Messages in $\phi$ and $\phi'$. \small  ${\mathbf B_c}$ denotes the
        sequence of encrypted ballot credentials in ${\mathbf B}$.
        ${\mathbf C_c}$ ($\tilde{\mathbf C_c}$) denotes the sequence of encrypted
        ballot credentials in ${\mathbf C}$ ($\tilde{\mathbf C}$), and
        ${\mathbf C_v}$ ($\tilde{\mathbf C_v}$) is the sequence of
        encrypted votes in ${\mathbf C}$ ($\tilde{\mathbf C}$).
        By $\hat b[x,y]$ and $\hat S_i$ we denote reencryptions of
        $b[x,y]$ and $S_i$ made by $\autt_0$.  $P_{u_0}$ ($\tilde
        P_{u_0}$) and $P_{w_0}$ ($\tilde P_{w_0}$) are the
        zero-knowledge proofs posted by $\autt_0$ in the mixing
        ballots and mixing credentials phase, respectively. 
        Messages that occur in $\phi$ and $\phi'$ at the same
        positions, are placed at corresponding positions in the table
        (for example $\hat S_3$ and $\hat S_1$).  } \end{figure*}
        %
    Only messages which are \emph{not produced} by the coercer (i.e.\
    neither constructed nor randomly generated by him,  like for
    instance messages posted by him on the bulletin board and
    forwarded to him back) are presented, as the messages produced by
    him  are not essential to the proof (instead of using them in
    $\tau_i$, one can use the corresponding terms that were used to
    construct them).  Also, we omit signatures on messages which are
    posted on the bulletin board. We only mention here, that the
    corresponding messages from left and right column, if signed, are
    signed by the some party (for instance, both messages in (1) are
    signed by $\autt_0$). We also omit the keys the coercer might
    have obtained form $\autk$ (these are the public keys of all the
    participants and the private keys of the dishonest ones).
    
    Messages (1) and (2) are posted by $\autt_0$ in steps (KGen1) and
    (KGen2). Messages (3) are posted by $\autr_0$ in step (Cred). (4)
    comprises messages sent by $\autr_0$ in step (Reg2) to dishonest
    voters who have requested for credentials. Note that up to this
    point messages in both $\phi$ and $\phi'$ are exactly the same.
    (5) contains the messages sent by $\autr_0$ to $\voter_0$ in step
    (Reg2) and forwarded to the coercer (in $\phi$) or a faked version
    of these messages (in $\phi'$).  (6) comprises votes posted by
    voters on ballot boxes.  Messages (7) are the commitments on the
    content of ballot boxes signed by these boxes and by the
    supervisor and posted on the bulletin board in steps (Comm1) and
    (Comm2).  (8) are PET shares and proofs posted by $\autt_0$ in the
    duplicate elimination phase of tabulation.  (9) and (10) are the
    results of mixing with reencryption of ballots posted by $\autt_0$
    in the mixing ballot phase and the mixing credential phase,
    respectively, along with the appropriate proofs.  By $\hat b[x,y]$
    and $\hat S_i$ we denote reencryptions of $b[x,y]$ and $S_i$ made
    by $\autt_0$. (11) contains the PET shares and proofs posted by
    $\autt_0$ in the invalid ballots elimination phase.  Finally, (12)
    are the distributed decryption shares and corresponding
    zero-knowledge proofs posted by $\autt_0$ in (Decr).

    One can check that condition (i) holds (i.e.\ that the
    $\chseq(\phi) = \chseq(\phi')$).  Hence, to complete the proof, it
    is enough to prove that the condition (ii) also holds, which is
    stated by the lemma bellow.

    \begin{lemma}\label{lem:civitas-steq} 
        For each $\tau_1, \tau_2 \in T_N$, we have that $\tau_1[\phi]
        \equiv \tau_2[\phi]$ iff $\tau_1[\phi'] \equiv \tau_2[\phi']$.
    \end{lemma}

    The remainder of this section is devoted to sketch the proof of
    this lemma.

    A \emph{destructor} is any of the following symbols:
    $\mathsf{first}$, $\mathsf{second}$, $\mathsf{unsig}$, 
    $\mathsf{checksig}$, $\mathsf{dec}$, $\distdec$, $\distpet$,
    $\public$, and $\checkznp$. The remaining symbols of
    $\Sigma$ are \emph{constructors}.  We will consider equations
    associated with destructors as rewriting rules, read from left to
    right (note that there is exactly one rule associated with each
    destructor).  Moreover, the equations associated with
    $\reenc(\cdot,\cdot,\cdot)$ will be also considered as rewriting rules.
    A term is said to be \emph{reduced}, if all the mentioned above
    equations, regarded as rewriting rules, are applied. 

    We will call $\phi$ and $\phi'$ \emph{frames}. We will sometimes
    write $\phi(x_i)$ instead of $x_i[\phi]$ (for the $i$-the element
    of $\phi$).  A frame is \emph{closed under applying destructors},
    if whenever a term of the form $g(x_i,t_1,\dots,t_n)[\phi]$, with
    some destructor $g$, reduces at the top (i.e.\ a reduction can be
    applied at the top level of the term), the result of total
    reduction of this term is also an element of the frame. We stress
    that such a result is a reduced term.
    
    Now, we define $\phi_0$ as a closure under applying destructors of
    $\phi$ and $\phi_1$ as the corresponding closure of $\phi'$, where
    ``corresponding'' means that the results obtained by applying the
    same terms (of the form $g(x_i,t_1,\dots,t_n)$) to both frames
    are, in both frames, at the same position. We will show that, for
    each $\tau_1, \tau_2 \in T_N$, we have that $\tau_1[\phi_0] \equiv
    \tau_2[\phi_0]$ iff $\tau_1[\phi_1] \equiv \tau_2[\phi_1]$,
    which immediately implies Lemma \ref{lem:civitas-steq}.
    
    A \emph{test} is an expression of the form $\tau_1 = \tau_2$.
    We will say that a test $\tau_1 = \tau_2$ holds in a frame $\phi$,
    if $\tau_1[\phi] \equiv \tau_2[\phi]$.  A test is \emph{basic}, if
    it is either of the form (a) $x_i = \tau$, where $\tau$ is a term
    with no destructor in the head, (b) $x_i = x_j$, or (c) $x_j =
    g(x_i,\tau_1,\dots,\tau_n)$, where $g$ is destructor and
    $\tau_1,\dots,\tau_n$ are some terms.

    We define the \emph{size} of a term in the usual way, but in case
    of terms representing zero-knowledge proofs ($ZK^{i,j}_\phi(\vec
    t)$), the size of the formula $\phi$ is taken into account too.

    The following lemma says that the frames $\phi_0$ and $\phi_1$ are
    indistinguishable w.r.t.\ basic tests.

    \begin{lemma}\label{lem:basic}
        For a basic test $\tau_1 = \tau_2$, the following is true:
        $\tau_1[\phi_0] = \tau_2[\phi_0]$ iff
        $\tau_1[\phi_1] = \tau_2[\phi_1]$. 
    \end{lemma}
    \begin{proof}[Sketch of Proof]
        (a) First, let us consider the case where the test is of the
        form $x = \tau$, where $x$ is one of $x_1,x_2,\dots$ and
        $\tau$ has no destructor in its head. We consider all $x$ case
        by case. For instance, let 
        $\phi_0(x) = \parenc{z_3}{\keyt}{r'_3}$  and
        $\phi_1(x) = \parenc{z}{\keyt}{r'_0}$ (these messages come
        from $b_3[z_3,s_3]$ and $b_0[z,s_0]$ posted by the voters on
        the bulletin boxes). Suppose that $x=\tau$ holds in $\phi_0$.
        $\tau$ cannot have a constructor in its head, because, there
        is no $\tau'$ such that $\tau'[\phi_0] \equiv r'_3$ ($r'_3$ is
        never revealed).  So, $\tau$ has to be a variable. However, no
        other variable gives a term equivalent to $\phi_i(x)$. Hence,
        $\tau$ must be $x$ and the test under consideration also holds
        in $\phi_{1-i}$.

        (b) If a test is of the form $x_i = x_j$, one can easily see,
        considering again case by case, that it holds in $\phi_1$ iff
        it holds in $\phi_2$.
        
        (c) If a test is of the form $x_j = g(x_i, \tau_1, \dots,
        \tau_n)$, where $g$ a is destructor and $\tau_1,\dots,\tau_n$
        are some terms, one should, again, consider all possible $x_i$
        case by case.  For instance, if $g = \distdec$ (i.e.\ a
        distributed decryption is applied), then $x_i$ must by a
        distributed decryption share provided by $T_0$ (Note that the
        destructor must reduce, because terms in $\phi_i$ are reduced
        and the test holds in one of $\phi_i$).  Hence, this
        decryption is applied to one of the encrypted ballots from the
        list $\mathbf C'$ of reencrypted and shuffled ballots, and the
        resulting votes are the same in both frames, and so the test
        does not distinguishes them.
    \end{proof}

    \begin{lemma}\label{lem:no-dest}
        Let $\tau_0 = \tau_1$ be a minimal test distinguishing
        $\phi_0$ and $\phi_1$. Then no destructor can be reduced in
        $\tau_j[\phi_i]$ ($i,j\in\{0,1\}$).
    \end{lemma}
    \begin{proof}[Sketch of Proof]
        For the sake of contradiction, let us suppose that some
        destructor can be reduced in $\tau_j[\phi_i]$. Let us consider
        a minimal subterm $\tau$ of $\tau_j$, with a destructor in its
        head, that can be reduced. Thus, its direct subterms are
        either variables of irreducible terms. If the left-most direct
        subterm of $\tau$ is a variable, then---because $\phi_i$ is closed under
        applying destructors---there is a variable $x_k$ such that
        $x_k = \tau_j$ holds in $\phi_i$. Now, by
        Lemma~\ref{lem:basic}, $x_k = \tau_j$ also holds in
        $\phi_{1-i}$. Hence the considered test is equivalent (in both
        frames) to $\tau_{1-j} = x_k$, and thus it is not minimal.

        If the left-most direct subterm of $\tau$, let us denote it
        by $\tau'$, is not a variable, then one can show that there is
        a subterm $\tau''$ of $\tau'$ such that $\tau[\phi_0] =
        \tau''[\phi_0]$ and $\tau[\phi_1] =
        \tau''[\phi_1\footnotesize]$, which is
        impossible, because $\tau^* = \tau_{1-j}$, where $\tau^*$ is
        obtained from $\tau_j$ by replacing $\tau$ by $\tau''$, would
        be a smaller test distinguishing the frames.  (We use here the
        observation that, in this case, a destructor can be applied
        only if some equations of some subterms of $\tau'$ hold in a
        frame, and because the considered test is assumed to be
        minimal, these equations must hold in both frames at the same
        time. So, in both frames the reduction can be applied.
        Moreover, in case, when the destructor in the head of $\tau_j$
        is $\distdec$ or $\distpet$, we use some particular properties
        of the frames under consideration and the fact that the
        arguments of these destructors can be freely rearranged.) 
    \end{proof}

    Finally, we prove the following fact which completes the proof of Lemma
    \ref{lem:civitas-steq}.
    \begin{lemma}
        For each $\tau_1, \tau_2 \in T_N$, we have that
        $\tau_1[\phi_1] \equiv \tau_2[\phi_1]$ iff $\tau_1[\phi_2]
        \equiv \tau_2[\phi_2]$
    \end{lemma}
    \begin{proof}[Sketch of Proof]
        For sake of contradiction, suppose that $\tau_0=\tau_1$ is a
        test which distinguishes these frames, i.e.\ it holds in
        $\phi_i$ and does not hold in $\phi_{1-i}$, for some
        $i\in\{0,1\}$. We can assume that
        this test is minimal (w.r.t.\ the size of terms).
        
        Assume that some of $\tau_j$, say $\tau_0$, is a variable.
        Then $\tau_1$ has to have a destructor in its head (because,
        otherwise, by Lemma \ref{lem:basic}, the test would not
        distinguish the frames). But, by Lemma \ref{lem:no-dest}, such
        a destructor cannot be reduced, so the test does not hold in
        neither of $\phi_i$ (since both frames are reduced and contain
        no destructors).

        Now, assume that none of $\tau_j$ is a variable.
        If we suppose that none of $\tau_0\phi_i$, $\tau_1\phi_i$
        reduces at the top, then one can construct a smaller
        test that distinguishes the frames, which contradicts the
        assumption about minimality of the test.
        Hence, it is enough to consider the case when some $\tau_j$,
        say $\tau_0$ reduces at the top position in $\phi_i$.

        By Lemma \ref{lem:no-dest}, we only need to consider three
        cases, depending of whether the top symbol of $\tau_0$ is (a)
        $\reenc$, (b) $+$, or (c) $\times$.  In each case, one obtains
        a contradiction. For instance, let us consider the case (a).
        So, $\tau_0\phi_i$ is of the form $\reenc(\reenc(m,k,r),k,r')$
        or $\reenc(\parenc mkr,k,r')$.  However, since $\tau_0$ is
        assumed to be reduced, $\tau_0$ has to be of the form
        $\reenc(x_l,k,r')$, for some variable $x_l$. Now, since the
        frames are reduced, $\phi_i(x_l)$ cannot be of the form
        $\reenc(m,k,r)$, so it must be of the form $\parenc mkr$.
        Hence, $\tau_0\phi_i$ is of the form $\reenc(\parenc
        mkr,k,r')$ and it reduces to $\parenc mk{r+r'}$.  Note that
        there is no term $\sigma$ such that $\sigma[\phi_{i}] \equiv
        r$, as $r$ is never revealed. It implies that $\tau_1$ has to
        be of the form $\reenc(x_{l'},k,r')$ with $\phi_{i}(x_{l'}) =
        \parenc mkr$ (there is no other way of obtaining $\parenc
        mk{r+r'}$). So, $x_l = x_{l'}$ holds in $\phi_i$ and, by Lemma
        \ref{lem:basic}, also holds in $\phi_{1-i}$.  It, however,
        means that the test $\tau_0 = \tau_1$ holds in
        $\phi_{1-i}$, which contradicts the assumption that it
        distinguishes the frames.
    \end{proof}



\section{Lee et al.\ Protocol}

    In this section we describe the protocol
    \cite{LeeBoydDawsonKimYangYoo-ICISC-2003} in more details and
    sketch the proof of coercion-resistance of this protocol.
    We can model cryptographic primitives used in this protocol, like
    in case of Civitas (see Fig. \ref{fig:eqth}), with some small
    modifications toward the \emph{threshold} decryption scheme. 

\subsection{Description of the Protocol}

    The set of agents we take is $\agents =  \{\voter_0, \dots,
    \voter_n, \auts, 
    {\Ttr}_{0}, \dots, \Ttr_n, $ $ \coercer,\autb,
    \autt_1,\dots,\autt_k,\autm_1, \dots,\autm_k\}$, where
    $\voter_0,\dots,\voter_n$ are voters, $\auts$ is the supervisor,
    $\Ttr_0,\dots,\Ttr_n$ are tamper-resistant randomisers, $\autb$ is
    the bulletin board, $\autt_1,\dots,\autt_k$ are the tallying
    authorities, $\autm_1,\dots,\autm_k$ are mixers, and $\coercer$ is
    the coercer. We assume that the coerced voter is $\voter_0$.
    
    All the messages posted on $\autb$ are publicly available.  The
    communication channel between a voter and his tamper-resistant
    randomiser is assumed to be untappable (i.e.\ it cannot be
    observed by the coercer). The remaining channels are public (can be
    observed by the coercer). 

    In the setup phase the tallying tellers $\autt_1,\dots,\autt_k$
    generate and publish his common public key $\keyt$ for the
    threshold decryption. Then, from the point of view of a voter
    $\voter_i$, the protocol execution consists of three steps:
        \begin{protocol}
            \mstep{P1}{\voter_i}{\Ttr_i}{ $ m_i $ }
            \mstep{P2}{\Ttr_i}{\voter_i}{ $ \sig{ m_i' }{k_{\Ttr_i}}$,}
            \stepcont{$\dvpr(\beta_i;\; m_i, m_i', \keyt, \pub(\voter_0)) $ }
            \mstep{P3}{\voter_i}{\aut}{ $  \sig{\sig{ m_i'
            }{k_{\Ttr_i} }}{\pub(k_{\voter_i})} $ }
        \end{protocol}
    where $m_i = \parenc{v_i}{\keyt}{\alpha_i}$, $m'_i =
    \reenc(m_i,\keyt,\beta_i)$, and $v_i$ denotes the vote
    chosen by $\voter_i$, $\alpha_i$ is a random value generated by this
    voter, and $\beta_i$ is a random value generated by $\Ttr_i$.  
    
    In the second phase of the protocol, the following steps are
    performed: (1) $\auts$ verifies the double signatures of voters
    and their randomisers on the posted ballots, and publishes valid
    ballots on the bulletin board.  (2) The, mixers
    $\autm_1,\dots,\autm_k$, in turn, shuffle and reencrypt these
    ballots, and post the result on the bulletin board.  (3) Talliers
    jointly decrypt shuffled ballots using the $(t,k)$-threshold
    ElGammal decryption protocol, and finally, (4) $\auts$ publishes
    the tally result.

    We assume that the correctness of all these steps is assured by
    posting appropriate non-interactive zero-knowledge proofs. This
    guarantees that only decryptions allowed by the protocol are
    performed, provided only a small fraction of the entities is
    dishonest.

\subsection{Proof of Theorem \ref{th:lee}}
    
    Recall that we want to prove coercion-resistance for the
    extended version of the protocol. The extension described in
    Section~\ref{sec:leeprotocol}, can be formalised as follows.
    The voter, instead of step (P1),
    performs the following step.
    \begin{protocol}
        \mstep{P1a}{\autt}{\autb}{ $m_i$, $P_i$ }
    \end{protocol}
    where
    $P_i$ is is a zero-knowledge proof which shows that the vote is
    well-formed with respect to the ballot design, i.e.\ $v_i$ is one
    of the valid votes (one can do it like in Civitas). Then,
    $\Ttr_i$, before replying with (P2), checks this proof.

    First, one can show that the protocol is normal and both $\alpha$
    and $\gamma$ are closed under $\peq$. Hence, we can use
    Theorem~\ref{th:dummy}. So, it is enough to provide a
    counter-strategy $v'$ for a strategy $v$ which simply forwards to
    the coercer all the messages obtained from the remaining
    participants and forwards to these participants all the messages
    obtained from the coercer.
    
    Let $z$ be a choice of $\voter_0$. Let $v'$ be the process which
    behaves like the forwarder $v$ with the following exception. When
    he is instructed to send a message $(m_c, P_c)$,
    then, instead, he sends ($m_0,P_0)$ as in specified in
    (P1a), and, instead of forwarding the answer of $\Ttr_0$ to the
    coercer, he sends him $m'_0$ signed by $\Ttr_0$ along with a
    faked DVRP for $m_c$ and $m_0'$.  We will show that $v'$ is a
    counter-strategy for $v$.

    \medskip \textbf{First, we show that condition (iii) of
    Definition~\ref{def:cr} holds} for $v'$. Let $\rho$ be a run of
    the system induced by $v'$, i.e.  $\rho$ is induced by $(v' \comp
    c \comp e)$, for some $c\in C, e\in E$.  If no message of the form
    $m^*$, as defined above, is posted on bulletin board and
    tallied, then there is nothing to prove.  So, suppose that
    some $\sig{\sig{m}{\Ttr_0}}{\voter_0}$ is posted and
    tallied. The only message signed by $\Ttr_0$ in $\rho$ is
    $m'_0$, which is a reencryption of the ballot containing the
    vote $z$, so $m=m'_0$.  Hence, as the tabulation phase has
    to be done correctly (because otherwise the authorities
    would not be able to construct valid zero-knowledge proofs),
    this vote is published.
    
    \medskip \textbf{Now, we will show that condition (i) of
    Definition~\ref{def:cr} holds} for $v$ and $v'$. So, let
    $\rho\in\alpha$ be a run induced by $(v \comp c \comp e)$, for
    some $c\in C$ and $e \in E$.
    Since $\rho$ is in $\alpha$, there is some honest voter, say
    $\voter_1$, who successfully votes for $z$. 
    
    We take $e'\in E$ which is as $e$ with the following exceptions:
    if, in $\rho$, a message of the form
    $\sig{\sig{m}{\Ttr_0}}{\voter_0}$ is posted on the bulletin board,
    where $m$ is a ballot with some vote $z_c$ (note that $z_c$ has to
    be a valid vote), then $\voter_1$, in $e'$, votes for $z_c$
    instead of $z$.

    The run $\rho'$ of $(v' \comp c \comp e')$ is constructed from
    $\rho$ in a natural way: the messages in $\rho'$ are
    delivered in the same order like the corresponding messages in
    $\rho$.  One can show that $\rho \sim \rho'$. The rough idea is as
    follows: $\voter_1$ is used to balance the outcome of the
    election.  Due to the fact that the coercer cannot tell any
    difference between an original DVRP and a faked one, and the fact
    that the messages posted on the bulletin board are mixed and
    reencrypted before decryption, the runs are indistinguishable to
    the coercer.

    \medskip \textbf{Now, we will show that condition (i) of
    Definition~\ref{def:cr} holds} for $v$ and $v'$. So, let
    $\rho\in\alpha$ be a run induced by $(v' \comp c \comp e)$, for
    some $c\in C$ and $e \in E$.  We define the vote of $z_c$: if the
    coercer voter, in $\rho$, is instructed to use his randomiser to
    reencrypt some ballot $m_c$ with a valid proof that $m_c$ contains
    a valid vote $v$, then $z_c$ is $v$; otherwise let $z_c$ by any
    vote.  Since $\rho$ is in $\alpha$, there is some honest voter,
    say $\voter_1$, who successfully votes for $z_c$. 
    
    We take $e'\in E$ which is as $e$ with the following exceptions:
    if, in $\rho$, a message of the form
    $\sig{\sig{m}{\Ttr_0}}{\voter_0}$ is post on the bulletin board
    (note that $m$ must be $m_c$ as defined above)
    then $\voter_1$, in $e'$, votes for
    $z$ instead of $z_c$.

    The run $\rho'$ of $(v' \comp c \comp e')$ is constructed from
    $\rho$, again, in a natural way: the messages in $\rho'$ are
    delivered in the same order like the corresponding messages in
    $\rho$.  One can show that $\rho \sim \rho'$, for the same reasons
    as previously.

\section{Okamoto Protocol} \label{sec:okamoto}

In this section we describe the protocol \cite{Okamoto-SPW-1997} and
discuss its properties.

\subsection{Cryptographic Primitives}

    In addition to the common cryptographic properties (which can be
    modelled like in Figure \ref{fig:eqth}), the protocol makes use of
    blind signatures and trapdoor commitment.  The equational theory
    associated with these primitives is given in
    Fig.~\ref{fig:eqth-okamoto}.
    \begin{figure}
        \begin{align*}
            \checksig{\sig{m}{k}}{\pub(k)} &= \True \\[1ex]
            \unblind{ \blsign{\blind{m}{t}{\pub(k)}}{k} }{t} &= 
                \sig{m}{k} \\[1ex]
            \tdcom{v'}{\f'(\alpha,v,r,v')}{\f(\alpha)} &= 
                \tdcom{v}{r}{\f(\alpha)} \\[1ex]
            \splitver{ \f(\parts{x_1,\dots,x_n}), \f(x_1), \dots, \f(x_n) }
                &= \True
            \end{align*}
        \caption{Equational Theory for Okamoto Protocol.\label{fig:eqth-okamoto}}
    \end{figure}

    These primitives are used in the following way. For a chosen vote
    $v$ and random values $r$ and $\alpha$, a voter can compute
    trapdoor-commitment for $v$, denoted by
    $\tdcom{v}{r}{\f(\alpha)}$. As it is only $\f(\alpha)$, not
    $\alpha$ itself, what is used to compute this expression, the
    commitment can be checked (recomputed) using $v$, $r$, and
    $\f(\alpha)$.  However, the voter, who also knows $\alpha$, can,
    for any vote $v'$, forge a value $r' = \f'(\alpha,v,r,v')$ which
    gives the same commitment value, i.e.\ $\tdcom{v}{r}{\f(\alpha)} =
    \tdcom{v'}{r'}{\f(\alpha)}$.

\subsection{Description of the Protocol}

    The set of agents is $\{\voter_0,\dots,\voter_n, \auta, \autb,
    \autt, \autr_1,\dots,\autr_N\}$, where $\voter_0,\dots,\voter_n$
    are voters, $\auta$ is an administrator, $\autb$ is a bulletin
    board, $\autt$ is a timeliness commission member, and
    $\autr_1,\dots, \autr_N$ are PRC members.

    Channels between $v_i$ and $\auta$ are network channels (the
    Internet). Messages posted on the bulletin board are sent trough an
    anonymous channel.  The voter $v_i$ send messages to $\autt$ and
    $\autr_j$ using untappable, anonymous channel.

    From the point of view of $\voter_i$, the protocol execution
    consists of the following steps: First, $v_i$ randomly generates
    $\alpha_i^1, \dots, \alpha_i^N$ and computes $\alpha_i =
    \parts{\alpha_i^1, \dots, \alpha_i^N}$.  Then he computes $G_i =
    \f(\alpha_i)$ and $G_i^j = \f(\alpha_i^j)$. Next, he randomly
    chooses $r_i$ and $t_i$.  Let
    \begin{align*}
        m_i &= \tdcom{v_i}{r_i}{G_i}, \\
        m'_i &= (m_i, G_i, G_i^1,\dots, G_i^N), \\
        x_i &= \blind{m'_i}{t_i}{\pub(\auta)}, \\
        z_i &= \sig{x_i}{\voter_i}
    \end{align*}
    Now, the following messages are exchanged:
    \begin{protocol}
        \mstep{P1}{\voter_i}{\auta}{ $\enc{x_i,z_i,\voter_i}{\pub(\auta)}$ }
        \mstep{P2}{\auta}{\voter_i}{ $y_i = \blsign{x_i}{\auta}$ }
    \end{protocol}
    As we mentioned, the communication channel between $v_i$ and $A$
    is a public channel.  Before executing (P2), $\auta$ checks the
    signature $z_i$ on $x_i$ and verifies that $\voter_i$ has the
    right to vote and he has not applied yet.  After step (P2) is
    performed, $\voter_i$ takes $s_i = \unblind{y_i}{t_i}$ (which is
    equivalent to $\sig{m'_i}{\auta}$). The successive steps are:
    \begin{protocol}
        \mstep{P3}{\voter_i}{\autb}{ $(m'_i, s_i)$ }
        \mstep{P4}{\voter_i}{\autt}{ $(v_i, r_i, m_i)$ }
        \mstep{P5}{\voter_i}{\autr_j}{ $(\alpha_i^j, G_i)$ }
        \mstep{P6}{\autr_j}{\autb}{ $u_i^j = (\f(\alpha_i^j), G_i), 
              \ \ \ \sig{u_i^j}{\autr_j}$}
    \end{protocol}

    In the counting stage, $\autt$, using messages from the bulletin
    board, checks whether the message obtained from $\voter_i$ in step
    (P4) is a valid ballot, as is explained below.  Then $\autt$
    publishes valid votes in random order.
    
    To check whether to accept a message (P4), $\autt$ does the
    following: He looks for the matching message $(m'_i, s_i)$ published on the
    bulletin board ($m'_i$ has to contain $m_i$ as the first
    component) and verifies that $s_i = \sig{m'_i}{\auta}$. Then he
    verifies that for $G_i$ taken from $m'_i$ it is true that $m_i$ is
    in fact equal to $\tdcom{b_i}{r_i}{G_i}$. He also checks whether,
    for all $G_i^j$ taken from $m'_i$, the corresponding message
    $(G_i^j,G_i)$ was published on the bulletin board by $\autr_j$ and
    that $\splitver{G_i, G_i^1,\dots,G_i^N} = \True$. If all these
    tests pass, the vote is accepted.  $\autt$ also provides a
    zero-knowledge proof that he has honestly published valid
    votes.

    \medskip

    In \cite{Okamoto-SPW-1997}, it is mentioned that the above voting scheme is not
    coercion-resistant, if one of the PRC's is not honest (cooperates
    with the coercer). So, a more complicated variant of the voting
    scheme (Scheme B) is also proposed. In this variant, the relation
    between $\alpha_i$ and $A = \{\alpha_i^1,...,\alpha_i^N\}$ is that
    it is enough to know some number $K<N$ of elements in $A$ to be
    able to compute $\alpha_i$ (as opposed to the variant presented
    above, where all $N$ elements of $A$ are necessary to compute
    $\alpha_i$).  

\subsection{Properties of the Protocol}

    In short, the Okamoto protocol does not provide coercion
    resistance even under strong assumptions. However, the
    protocol is interesting in that it highlights the difference
    between single-voter coercion and multi-voter coercion, in
    absence of dishonest voters.
    
    In \cite{Okamoto-SPW-1997}, the proof of coercion-resistance
    is based on the observation that the only way to make a
    ballot accepted is to send valid $\alpha_i^j$ to all
    $\autr_j$. So because the channels between the voter and
    PRC's are untappable, it can be only the voter who sends
    these values and, in consequence, he has to know them.
    Thus, he is able to compute $\alpha_i$ and make up a value
    $r'$ which enables him to vote for the vote of his choice.
    This reasoning misses, however, the fact that different
    $\alpha_i^j$ can be sent by different voters or even by the
    coercer, if he is also a voter.

    If we assume that the coercer is an entitled voter or there
    is some dishonest voter, the protocol is clearly not
    coercion-resistance: The coercer prepares a ballot, ask the
    coerced voter to obtain a blind signature on this ballot and
    then completes the process by himself, using the anonymous
    untappable channels he has access to.

    \enlargethispage{-6.7cm}
    Event if we assume that the coercer is not an entitled voter and
    there is no dishonest voter, then still the protocol is not
    coercion-resistant, provided that more than one voter is
    coerced at the same time.  In this case the coercion
    strategy is as follows. All the coerced voters are supposed
    to obtain a blind signature of the appropriate voting
    authority on messages provided by the coercer.  Then, the
    coercer distributes the private credential shares to the
    voters in such a way that no coerced voter has a complete
    collection of private credential shares, i.e., the shares
    for one vote are distributed among different coerced voters.
    As a result, no coerced voter can open his/her commitment in
    an arbitrary way. This suffices for the ballots of the
    coercer to be accepted.  To the best of our knowledge, this
    attack has not been observed before.  A more detailed
    description of this attack follows.
    
    Suppose that there are $N$ voters that are coerced (recall
    that $N$ is the number of PRC's; we chose this number for
    simplicity of the proof).  The coercion strategy is as
    follows.  All the coerced voters are supposed to obtain a
    signature on messages provided by the coercer.  The $i$-th
    message is build, as $m'_i$ in the protocol description,
    using $\alpha_i^1,\dots,\alpha_i^N$.  Each $\voter_i$ is
    then supposed to send $x_i$, like in the protocol
    description, for $v_i$ chosen by the coercer.  Furthermore,
    $\voter_i$ is supposed, for each $j\in\{1,\dots,N\}$, to
    forward $\alpha_{a}^j$ to $\autr_j$, where $a=(i+j)\mod N$.
    These shares are the only ones that the voter learns. So, he
    is not able to compute any of $\alpha_i$ (it is also true in
    scheme B), because he knows only one private share for each
    $\alpha_i$. Thus, the only valid vote $\voter_i$ can send to
    $\autt$ is $v_i$, as demanded by the coercer. Because
    $\autt$ provides a zero-knowledge proof that the submitted
    votes are accounted for, the coercer can verify, that this
    vote has been really posted by the voter.

    While the above attack allows the coercer to vote as he
    wishes, an abstention attack is possible even if only one
    voter is coerced, the coercer is not entitled to vote and
    there are not dishonest voters. 

    The only setting in which we could prove coercion
    resistance of the Okamoto protocol is in the setting just
    described where $\alpha$ is defined similarly to the Lee et
    al.~protocol and the goal $\gamma$ is merely that
    if the coerced voter posts message (P3) on the
    bulletin board, then his/her successfully votes for the
    candidate of his/her choice.
    
    Note that for this result to hold it is essential that only
    one voter is coerced.



\end{document}